# Minimizing round-trip loss for near-unity quantum efficiency in organic light-emitting diodes

Junho Kim[1,2], Seonjeong Lee[3], Seunghyup Yoo[2,3], Malte C. Gather[1,4]

[1] Humboldt Centre for Nano- and Biophotonics, Department of Chemistry and Biochemistry, University of Cologne, Köln, Germany.

[2] School of Electrical Engineering, Korea Advanced Institute of Science and Technology (KAIST), Daejeon, Republic of Korea.

[3] Graduate School of Semiconductor Technology, Korea Advanced Institute of Science and Technology (KAIST), Daejeon, Republic of Korea.

[4] School of Physics and Astronomy, University of St Andrews, North Haugh, St Andrews, UK.

Correspondence: Seunghyup Yoo (syoo_ee@kaist.ac.kr), Malte C. Gather (malte.gather@uni-koeln.de)

## Abstract

Modern organic light-emitting diodes (OLEDs) reach internal quantum efficiencies close to 100%[1,2], yet in a planar device only 20–30% of photons escape[3-5]. External outcoupling structures on high-index substrates recover some of the trapped light, but external quantum efficiencies (EQEs) have stalled at 55–60% across different optical designs[6,7], and the origin of this ceiling is not fully understood. Here we identify light extraction in such OLEDs as a multi-pass process: light that the outcoupling structure fails to extract returns to the OLED, is reflected back and may escape over multiple passes, with the efficiency governed by the absorption per round trip. Absorption loss is minor in conventional planar OLEDs, but it becomes dominant and accounts for the observed efficiency ceiling when an outcoupling structure is attached. Consistent with this picture, adding a low-loss auxiliary reflector raised the EQE of an OLED on a high-index microlens substrate from 51% to 77% without any change to the OLED stack or the outcoupling structure, the highest value reported for a single-junction OLED without a macroscopic extraction lens; for an OLED on a conventional glass substrate an estimated 92% of the light delivered to the substrate was extracted. We derive four design rules that keep round-trip losses to a few percent and predict that EQEs around 90% are feasible with established materials. Under these rules, microcavity tuning and emitter orientation largely cease to matter, and high efficiency is retained across a wide range of microlens designs and scattering layers.

## Introduction

Organic light-emitting diodes (OLEDs) lead today's display market owing to their color purity, surface emission and unsurpassed compatibility with large-area processing[8], and are spreading to augmented- and virtual-reality devices[9] and to wearable and implantable light sources[10]. Following the development of phosphorescent[1] and thermally activated delayed fluorescence[2] emitters, their internal quantum efficiency is now close to 100%. However, the external quantum efficiency (EQE), the ratio of the number of photons emitted into air to the number of injected electrons, remains much lower because the refractive-index contrast between the organic layers, the substrate and air traps most light inside the device, with only 20–30% escaping from conventional planar OLEDs[3-5]. Therefore, OLEDs generally remain less efficient than inorganic LEDs and are not yet considered able to replace them where high brightness and long lifetime are required[11].

Recovering trapped light is a main route to further improving OLED efficiency. Emitters with horizontally oriented transition dipoles[12-14] and microcavities[15,16] direct more of the emission into the escape cone, whereas external outcoupling structures, such as microlens arrays and scattering layers[6,7,17,18], extract light trapped in the substrate without interfering with the electrical operation of the device. Combined with a high-index substrate or a low-index transparent electrode, external outcoupling structures also give access to light that would otherwise be waveguided in the OLED layers. Many different optical designs have been proposed, some adding considerable complexity to the device, yet the EQE of OLEDs has generally plateaued at 55–60%[6,7]. This convergence suggests a common limiting mechanism rather than a shortcoming of any particular structure. Higher EQE values have been obtained only when the substrate light is collected by a macroscopic lens[19,20], and this shortfall has been noted as a general bottleneck[21]. However, the underlying loss channels have not been fully quantified, there are no guidelines for reliably exceeding 60%, and the attainable limit remains unknown.

Here we show that outcoupling is limited by the OLED itself, not by the outcoupling structure. Most outcoupling structures extract only a fraction of the light that reaches them, a fact that has received little attention. The remainder is returned to the device, where it is partly absorbed and partly reflected by the layer stack and sent back for another attempt. Extraction therefore takes place over many passes, and its efficiency ultimately depends on how much light the device absorbs per round trip. Unlike most earlier work, which aimed to make the outcoupling structure extract as much light as possible on its first pass, we focus on minimizing this round-trip absorption.

We capture the multi-pass process in a closed-form expression with two parameters, the escape probability per pass and the round-trip loss, which shows that losses that are of little importance in conventional OLEDs are amplified by the repeated passes. A simulation that couples wave optics in the thin-film stack of the OLED to ray optics in the substrate supports this picture. It shows that because microlens arrays return light largely at its original angle instead of randomizing it, real devices generally need even more passes than the closed-form expression assumes, which tightens the requirement on loss further. In experimental tests, devices built to minimize round-trip losses extracted up to an estimated 92% of the light delivered to the substrate; adding an Al auxiliary reflector to a bottom-emitting OLED on a high-index microlens-array substrate increased its EQE from 51% to 61%, and replacing the Al reflector with a low-loss reflector raised it further to 77%.

Finally, we identify a set of design rules to further increase light-extraction efficiency in the future. Calculations following these rules indicate that the EQE can be enhanced to well above 80% even with an isotropically oriented emitter and to 90% for a low-absorption transparent electrode. Our approach leaves a wide design space and should extend beyond conventional bottom-emitting OLEDs to top-emitting OLEDs, tandem devices, quantum-dot LEDs and perovskite LEDs.

## Results

### Light extraction as a multi-pass process

Figure 1a compares a conventional OLED with one designed for minimum round-trip loss, each fitted with an outcoupling structure. Light that is not extracted on its first encounter with the outcoupling structure returns to the device, is reflected by the layer stack and reaches the outcoupling structure again. Because extraction takes place over many passes, even a small loss per round trip, e.g., from Joule dissipation in the metal electrode or from absorption in the transparent conductive oxide (TCO) electrode, strongly reduces the final efficiency. By contrast, when these losses are suppressed, most of the light delivered to the substrate is eventually extracted even if the fraction of light extracted during each encounter with the outcoupling structure is small.

To separate the losses inside the device from those of the extraction process, we write the EQE as the product of two factors:

$$\mathrm{EQE} = \eta_{\mathrm{sub}} \times \eta_{\mathrm{sta}} \qquad (1)$$

The substrate-delivery efficiency $\eta_{\mathrm{sub}}$ is the fraction of light generated in the emitting layer that reaches the substrate after losses to waveguided modes, surface plasmon polariton (SPP) modes and immediate absorption in the device. The substrate-to-air extraction efficiency $\eta_{\mathrm{sta}}$ is the fraction of the light in the substrate that is emitted into the surrounding air. Charge balance and exciton utilization are taken as unity, and the photoluminescence quantum yield (PLQY) of the emitter is included in $\eta_{\mathrm{sub}}$ (Supplementary Note 1).

In a conventional planar device, $\eta_{\mathrm{sta}}$ is fixed by the escape cone[5] and scales as $1/n_{\mathrm{sub}}^2$ for isotropic light in the substrate. With an outcoupling structure, $\eta_{\mathrm{sta}}$ is instead set by the balance between light extraction and loss during each round trip. Let *p* be the probability that light escapes into air in a single encounter with the outcoupling structure, $R_{\mathrm{LED}}$ the reflectance of the device for light incident from the substrate, and $A' = 1 - R_{\mathrm{LED}}$ the fraction lost per round trip (by absorption and, for a partly transparent device, transmission through the rear). If the outcoupling structure does not absorb and randomizes the angle of the light on every pass, the fractions extracted in successive round trips form a geometric series with ratio $(1-p)(1-A')$, which sums to

$$\eta_{\mathrm{sta}} = \frac{p}{p+(1-p)A\prime} \qquad (2)$$

For a device that transmits no light through the rear, this expression is formally identical to the statistical-optics result for light extraction from thin-film inorganic LEDs with a rear reflector[22-24]. However, for a thin-film OLED, $A'$ is the parasitic absorption of the multilayer stack, set mainly by the reflector and the transparent electrode, and *p* is of order $1/n_{\mathrm{sub}}^2$ for light with randomized angles, i.e., about 0.4 for a glass substrate ($n_{\mathrm{sub}} = 1.5$)[23].

Equation (2) illustrates why small losses matter. For *p* = 0.4, round-trip losses of $A' = 0.02$, 0.10 and 0.15 limit the achievable cumulative extraction to 0.97, 0.87 and 0.82, respectively (Fig. 1b); that is, for a large loss, most light is absorbed within a few round trips, whereas it continues to recirculate until nearly all of it has escaped when $A'$ is small. Consequently, $\eta_{\mathrm{sta}}$ rises steeply as the round-trip reflectance approaches one (Fig. 1c, for *p* = 0.25, 0.4 and 0.6), and near-unity extraction becomes possible when the parasitic absorption is kept to a few percent.

The assumption of full angular randomization is not met by many real outcoupling structures. For instance, a close-packed hemispherical microlens array returns light largely at its original angle (Supplementary Fig. 1) and transmits little light that arrives at large angles of incidence (Supplementary Fig. 2), so light remaining after the first pass escapes less readily than equation (2) predicts. We therefore also evaluate extraction with an angle- and wavelength-resolved model, in which the angular distribution of light is propagated from pass to pass by matrix multiplication (Methods and Supplementary Note 1). For microlens arrays and the device stacks in Supplementary Table 1, this gives an $\eta_{\mathrm{sta}}$ that is 1.6–3.2 percentage points lower than equation (2) predicts; larger deviations occur close to the cut-off of waveguided modes (Methods and Supplementary Fig. 3). Deviations from the idealized full angular randomization behavior thus make a low round-trip loss even more important. In the following, we use equation (2) to establish design trends and the angle-resolved model to compute the efficiency of specific devices.

## Origin of the round-trip loss

In conventional OLEDs, absorption in the transparent electrode and reflection loss at the metallic electrode (Joule dissipation) are secondary losses, and the thickness of the transparent electrode is typically determined mostly by the need to satisfy the cavity resonance condition[4,25]. With an outcoupling structure, however, electrode absorption and reflection loss occur during every pass, substantially reducing $\eta_{\mathrm{sta}}$. Figure 2a–c shows the round-trip loss as a function of transparent electrode thickness for an ITO-based OLED, separated into Joule dissipation at the reflective metal electrode and absorption in the ITO. With an Al electrode, the reflection loss alone amounts to 13% of the light per round trip, and a 150-nm-thick ITO electrode accounts for only about 15% of the total loss. When the metallic electrode material is replaced by Ag, which is the least absorbing metal at visible wavelengths, the reflection loss falls to 1.8% per round trip, and a 150-nm-thick ITO electrode then becomes the dominant loss channel. Consequently, $\eta_{\mathrm{sta}}$ is 11–13 percentage points higher for an Ag electrode than for Al; it decreases from 97% to 93% (Ag electrode) and from 84% to 82% (Al) as the ITO thickness increases from 30 to 200 nm (Fig. 2c).

The penalty for round-trip loss grows with the refractive index of the substrate and outcoupling structure (Fig. 2d–f). A higher index reduces the critical angle and lowers $p$, so light needs more passes to escape (Supplementary Figs 1 and 2). Going from $n_{\mathrm{sub}} = 1.3$ to 2.0 reduces $p$ from 0.58 to 0.23 (Fig. 2d). A high-index substrate thus has two opposing effects: it lowers $\eta_{\mathrm{sta}}$ but raises $\eta_{\mathrm{sub}}$ (Fig. 2e,f). In an OLED using a conventional Al electrode, the loss in $\eta_{\mathrm{sta}}$ offsets much of the gain in $\eta_{\mathrm{sub}}$: the achievable EQE reaches a maximum of around 60% for $n_{\mathrm{sub}} = 1.8$ and then slowly decays for larger $n_{\mathrm{sub}}$ (Fig. 2f). The high round-trip loss caused by the conventional Al electrode thus explains the EQE ceiling observed in the current OLED literature[6,7]. In a device with a lower-loss Ag electrode, by contrast, the expected EQE rises to 89% at $n_{\mathrm{sub}} = 1.8$ and then levels off. The angle-resolved model reproduces both EQE levels, although it reaches them at a slightly higher substrate index (Supplementary Fig. 3d).

The angle and wavelength dependence of the round-trip loss varies strongly with the choice of material for the reflective electrode (Fig. 2g–j). For a device with a 150-nm-thick transparent ITO electrode (see Methods for full OLED stack), the loss weighted over the emission spectrum and over the angular distribution of the light in the substrate ($\cos\theta \sin\theta$) is 15.6% for a conventional Al cathode and only 4.5% for Ag. Because the light delivered to the substrate and the light returned by the outcoupling structure each have their own angular distribution, what matters for a given electrode is that the absorption is low at the angles that carry most of the power. A Poynting-vector calculation shows the contributions of the ITO and the metal electrode to the overall loss (Supplementary Fig. 4).

Joule dissipation at the reflective metal electrode can be avoided if the metal is replaced with a dielectric distributed Bragg reflector (DBR) placed behind a second transparent electrode. We designed a DBR comprising ten thermally evaporable ZnS/LiF pairs as a chirped stack that reflects over all escape angles and the whole emission spectrum (Fig. 2i,j; Supplementary Note 2 and Supplementary Fig. 5). In an OLED stack where this DBR is placed behind a 50-nm-thick ITO cathode, the total round-trip loss is 5.1%, of which 4.1% is due to absorption in the ITO anode and cathode and only 1.0% originates from transmission through the DBR outside its reflectance band. This value is comparable to the 4.5% round-trip loss obtained with a simpler design based on an Ag electrode. However, as a DBR needs no electrical contact to the OLED stack, it can be added to a finished device as an auxiliary reflector, an option we use in the experiments below.

## Experimental test

To test whether lowering the round-trip loss raises the extraction efficiency as equation (2) predicts, we fabricated two types of phosphorescent OLEDs, using microlens arrays as external outcoupling structures in both cases (Fig. 3). In OLEDs with a small active area and a thick substrate, the light returned by the outcoupling structure spreads laterally, and much of it reaches the rear of the device outside the active area, where there is no reflective electrode. We therefore added auxiliary reflectors that are much larger than the active area to return this light to the outcoupling structure (Methods and Supplementary Fig. 6). Because all auxiliary reflectors lie outside the path of electrical current, they change only the round-trip loss and leave the electrical operation of the device unaffected.

The first device is a green phosphorescent OLED with a low-loss 100-nm-thick Ag electrode on a conventional glass substrate, to which a commercial microlens-array film was attached (Fig. 3a). An electrically isolated Ag film deposited on top of the thin-film encapsulation of the device serves as auxiliary reflector. The maximum EQE increased 2.8-fold, from 20% for the planar reference to 55.8%, in agreement with the 55.9% predicted by our simulation; multiple devices fabricated in the same batch showed nearly identical characteristics (Supplementary Fig. 7). Based on the $\eta_{\mathrm{sub}}$ of 60.9% that was calculated for this device stack, the substrate-to-air extraction efficiency reaches $\eta_{\mathrm{sta}}$ = 91.6% in this OLED (Methods).

The second device combines a low-loss auxiliary reflector with a thin transparent electrode and a high-index substrate. It is based on an orange-emitting phosphorescent OLED with a conventional Al cathode and a low-loss 50-nm-thick IZO anode deposited onto a high-index substrate with a microlens array ($n_{\mathrm{sub}} = 1.77$; Fig. 3b). Without an auxiliary reflector, this device reached an EQE of 51%, compared with 23% for the planar reference. With an Al mirror placed behind the device as auxiliary reflector, the EQE rose to 61%, consistent with the limit of around 60% calculated for an Al reflector (Fig. 2f). Depositing a DBR consisting of 4.5 ZnS/LiF pairs on top of a device with the same stack instead raised the EQE to 77%, which corresponds to $\eta_{\mathrm{sta}}$ = 87.5% using the $\eta_{\mathrm{sub}}$ of 88% calculated for this device. The DBR substantially reduces the fraction of the generated light that leaks out through the back of the device, which strongly improves device efficiency (Fig. 3c and Supplementary Fig. 8). We note that the DBR used here is a uniform stack and simpler than the chirped design of Fig. 2i (Supplementary Note 2). Nevertheless, this result shows that a low-loss auxiliary reflector can raise extraction efficiency even in the presence of a lossy Al cathode, as expected from the multi-pass model, without any change to the OLED stack itself or to the external outcoupling structure. To our knowledge, the EQE of 77% is the highest reported for a single-junction OLED with a microscopic outcoupling structure; it exceeds the 55–60% of earlier such devices and is comparable to the highest values previously reached with macroscopic lenses or index-matched detection[19-21].

## Attainable efficiency and design freedom

With a low-loss reflective electrode in place, absorption in the transparent electrode is the main remaining loss in a bottom-emitting OLED. To minimize this loss, we propose to make the transparent anode as thin as reliable electrical operation allows (e.g., 50 nm of ITO or 10 nm of Ag), instead of tuning its thickness for cavity resonance. Figure 4a shows how sensitive the total efficiency of an OLED with a low-loss reflector is to the optical constants of the transparent anode. For ITO, where the extinction coefficient *k* is sensitive to the process conditions during deposition[26,27] and varies even through the thickness of a single film[28], $\eta_{\mathrm{sub}}$ decreases almost linearly with *k*, whereas $\eta_{\mathrm{sta}}$ falls hyperbolically, as expected from equation (2). Raising *k* from 0 to 0.01 lowers $\eta_{\mathrm{sta}}$ from 95% to 81% and thus reduces the attainable EQE from 91% to 75%. The extinction coefficient of ITO is therefore a decisive material parameter. For thin metal anodes, the corresponding quantity is the real part of the refractive index *n*: alloying Ag with other metals to obtain continuous ultra-thin films[29,30] raises *n* to 0.2 or more and thereby limits the achievable EQE to 55% or less. The same consideration applies to top-emitting OLEDs, in which the transparent top electrode is usually a thin metal film: these structures will likely require a continuous thin Ag layer grown with seed-layer control, or a low-damage sputtered oxide[31].

These considerations lead to two important design rules: light delivered to the substrate escapes efficiently if the rear reflector has a round-trip loss of no more than a few percent (Rule 1) and the transparent electrode is as thin and as weakly absorbing as its electrical function allows, instead of having its thickness set by the cavity resonance condition (Rule 2). Two further aspects are important to maximize $\eta_{\mathrm{sub}}$: matching the substrate index to that of the emitting layer moves the waveguided modes into the substrate[32,33] (Rule 3, Supplementary Fig. 9); thin transparent electrodes help here too, because they are too thin to guide light by themselves. SPP loss is reduced by a thick electron-transport layer (ETL) or by an ETL with a low extraordinary refractive index (Rule 4)[4,34]: the ETL thickness required for an evanescent loss below 20% falls from 110 nm to 59 nm as the index is lowered from 1.8 to 1.5 (Supplementary Fig. 10). These four rules act in concert; for instance, while using a high-index substrate increases the fraction of light delivered into the substrate, it lowers the probability of light extraction on each pass, so its benefit is realized only when the round-trip loss is small.

In a conventional OLED, the thicknesses of the electron- and hole-transport layers are adjusted to tune the device to an optical resonance, and the calculated EQE shows pronounced interference fringes as the transport-layer thickness is changed (Fig. 4b). In an OLED that follows the design rules laid out above, these fringes disappear, and our calculations indicate that more than 90% of the maximum achievable EQE is obtained for any ETL thickness above 150 nm (Fig. 4c). Emitter orientation also becomes less important. The EQE of a conventional OLED with an Al cathode on glass and a microlens array rises from 31% to 55% as the horizontal dipole ratio increases from 0.5 to 1.0, which is one reason why much emitter research aims at achieving preferential horizontal orientation[12,13,35,36]. However, with a low-loss OLED stack and an index-matched substrate, the EQE varies by only two percentage points over the same range (Fig. 4d). Using the measured optical constants of a representative OLED stack, the angle-resolved model predicts EQEs of 87 to 89% for isotropic and oriented emitters alike, and EQEs up to 91% are predicted if the extinction coefficient of ITO is lowered from $k = 0.0032$ to 0.002 (Supplementary Table 1). The exact design of the outcoupling structure used also becomes less important: the EQE stays above 95% of its maximum for microlens aspect ratios from 0.3 to 1.5, and when using scattering layers instead of microlenses, it remains at 80–90% over a wide range of scattering strength and anisotropy (Supplementary Figs 11 and 12). This reduces practical burdens such as the optimization of the process for making the scattering layer as well as challenges associated with sample-to-sample variation.

## Discussion

Describing light extraction from OLEDs with an external outcoupling structure as a multi-pass process changes the view of what limits their efficiency. The absorption per round trip, a loss of little consequence in conventional devices, accounts for the ceiling near 60% that different optical designs have run into, and reducing it becomes a direct handle for raising the efficiency. Four design rules follow, two that minimize this loss and two that maximize the light delivered to the substrate. In experiments, low-loss auxiliary reflectors gave $\eta_{\mathrm{sta}} = 91.6\%$ on glass and raised the EQE of an OLED on a high-index microlens substrate to 77%. Calculations using the experimentally determined optical constants of established OLED materials indicate that EQEs around 90% are within reach. Because the extraction efficiency depends only on the magnitude of the round-trip loss and not on how it is achieved, metallic and dielectric reflectors can be used alike. The rules also remove the need for microcavity tuning and emitter orientation and make the choice of outcoupling structure largely uncritical. With internal quantum efficiencies now close to unity thanks to decades of progress in emitter and stack design, the losses of the reflector and the transparent electrode emerge as new key parameters for improving OLED efficiency.

So far, our calculations have treated the organic layers as non-absorbing; any residual absorption in them adds to the round-trip loss and in future work this will need to be quantified, particularly for the thick transport layers used to suppress SPP losses. The auxiliary reflectors used here require space outside the active area. In top-emitting display pixels, the low-loss reflector has to be integrated into the device itself, for instance, by using a low-loss reflective bottom electrode and a transparent top electrode with minimal absorption, such as an ultra-thin Ag electrode or a low-damage sputtered oxide electrode. The expected increases in EQE can reduce the electrical power required for a given light output by up to one third, which bears directly on the brightness and lifetime limitations that have so far favored inorganic LEDs in certain applications. Finally, as the underlying principle of our work is not specific to organic emitters, it should apply equally to quantum-dot and perovskite LEDs and may serve as a general basis for the optical design of thin-film light-emitting devices.

# Methods

## Optical simulation

The optical simulations combine ray optics, wave optics and an integrated model that couples the two. The bidirectional scattering distribution function (BSDF) of the outcoupling structure was computed using Monte Carlo ray tracing (LightTools) with at least 5,000,000 rays per angle of incidence at a wavelength of 550 nm. The BSDF was assumed to be constant across the visible range. All wave-optical calculations were performed with custom MATLAB code. The absorption, reflectance and transmittance of the multilayer stack were obtained with the transfer-matrix method, the emission characteristics of the OLED were described with the classical dipole model[37,38], and the power dissipation within the thin films was obtained with a Poynting-vector model[39]. The optical constants of ITO[40] and Ag[41], the birefringence of the organic layers (measured ordinary and extraordinary refractive indices)[42], the PLQY, the horizontal dipole ratio and the Purcell factor were included, and the emission zone was placed at the center of the EML. The trans-scale optical simulation follows ref. [43]; in brief, the emission profile delivered to the substrate is computed for a semi-infinite substrate and the reflectance of the device for light incident from the substrate is taken into account.

For the angle- and wavelength-resolved form of equations (1) and (2), we write the power delivered to the substrate as $P_0$ and its normalized angular distribution as the row vector $P_{\mathrm{sub}}(\theta)$, the transmitted and reflected components of the outcoupling structure as the row vector $B_{\mathrm{T}}(\theta)$ and the matrix $B_{\mathrm{R}}(\theta_i, \theta_r)$, and the reflectance of the device as the diagonal matrix $R(\theta)$. The extracted power summed over all orders of round trips is then $P_{\mathrm{out}} = P_0 \cdot P_{\mathrm{sub}} \cdot (\mathrm{I} - B_{\mathrm{R}}R)^{-1} \cdot B_{\mathrm{T}}^{T}$. The series was summed until convergence (60–100 round trips unless stated otherwise), evaluated at each wavelength and weighted by the emission spectrum of the emitter used. Replacing the angle-dependent quantities by their averages over the recirculating light, $p$ and $A' \equiv 1 - \langle R \rangle$, and using $B_{\mathrm{R}} \approx 1 - B_{\mathrm{T}}$ reduces the series to equation (2) (Supplementary Note 1). The escape probability $p$ is the transmitted component $B_{\mathrm{T}}(\theta)$ of the BSDF averaged with the angular weight $\cos\theta \sin\theta$, and is a property of the outcoupling structure alone. The BSDFs are ray-tracing results for hexagonally close-packed hemispheres with the refractive index of the hemispheres $n_{\mathrm{MLA}}$ equal to the substrate index. Because the series uses $P_{\mathrm{sub}}(\theta)$ without averaging, it can differ from equation (2) by up to 9 (Ag) and 12 (Al) percentage points just above each index at which a waveguided mode enters the substrate at a near-grazing angle ($n_{\mathrm{sub}}$ ≈ 1.42–1.46, 1.58–1.59 and 1.75–1.78 at 550 nm, depending on the cathode, and at $n_{\mathrm{sub}}$ = $n_{\mathrm{org}}$ = 1.8). This is mainly because $P_{\mathrm{sub}}(\theta)$ is not Lambertian and the outcoupling structure filters by angle (Supplementary Note 1 and Supplementary Fig. 3). The positions of these dips depend on the mode structure of the specific stack, and the EQE shows no corresponding dips because the light that causes the dips also raises the substrate-delivery efficiency (Fig. 2f and Supplementary Fig. 3d). For Fig. 2c–f, which shows design trends independent of the materials and structure, we use equation (2).

The design-rule calculations (Figs 2a–f, 4a,d and Supplementary Figs 9–11) were performed at 550 nm with emitter PLQY = 1 and isotropic emitter orientation ($\Theta$ = 2/3; $\Theta$ is varied only in Fig. 4d), and with isotropic, non-absorbing charge-transport and emitting layers ($n$ = 1.8); in Supplementary Fig. 10 only, the ETL is birefringent ($n_{\mathrm{o}}$ = 1.8, $n_{\mathrm{e}}$ varied). The stack of Fig. 2 is substrate ($n_{\mathrm{sub}}$ = 1.5, varied in Fig. 2d–f)/ITO (varied in Fig. 2a–c, 50 nm in Fig. 2d–f and 150 nm in Fig. 2g–j)/420 nm of organic layers (emitter at the center)/100-nm reflector. Supplementary Fig. 11 uses the stack of the proposed device in Fig. 4d (reflector 100 nm/ETL 200 nm/EML 20 nm/HTL 200 nm/ITO 50 nm/substrate with $n_{\mathrm{sub}}$ = 1.8); lenses with aspect ratios above 1 are half-ellipsoids. Figure 4b,c was computed for the stack reflector 100 nm/B3PyMPM $d_{\mathrm{ETL}}$/TCTA:B3PyMPM 25 nm (emission zone at the center)/TAPC $d_{\mathrm{HTL}}$/ITO 50 nm/substrate with ETL and HTL thicknesses of 10–300 nm, for (b) an Al reflector on a planar substrate ($n_{\mathrm{sub}}$ = 1.5) and (c) an Ag reflector with a hemispherical microlens array ($n_{\mathrm{sub}}$ = $n_{\mathrm{MLA}}$ = 1.8, 20 terms of the series), over 400–700 nm in 1-nm steps weighted by the emission spectrum of $Ir(ppy)_2acac$ on a photon-number basis. The in-plane wavevector was sampled with 1,000 points per unit up to three times the emitting-layer wavevector. For the index-matched, non-absorbing scattering layer in Supplementary Fig. 12, the BSDF was generated by Monte Carlo ray tracing ($5\times10^4$–$10^5$ photons per incidence-angle bin) with a Henyey–Greenstein phase function[7,44] ($g$ = 0.5–1.0) and a scatterance $S = \mu_s d$ = 1–15, for the device reflector (Ag or Al) 100 nm/B3PyMPM 500 nm/TCTA:B3PyMPM:Ir(ppy)$_2$acac 25 nm (emission zone at the center)/TAPC 500 nm/ITO 50 nm/substrate ($n_{\mathrm{sub}}$ = 1.8); its EQE was obtained by summing 20 round trips over 400–800 nm (1-nm steps) with PLQY = 0.98, $\Theta$ = 0.76. For Supplementary Table 1, $P_{\mathrm{sub}}$ and $R_{\mathrm{LED}}$ were obtained at 550 nm and over 400–700 nm, and the series was summed with the same microlens-array BSDF ($n_{\mathrm{MLA}}$ = 1.8) as in Fig. 4.

## Definition of the extraction efficiency

For the devices of Fig. 3, $\eta_{\mathrm{sta}}$ was obtained by dividing the measured EQE by the modeled $\eta_{\mathrm{sub}}$ (0.609 for the green and 0.880 for the orange device) and thus combines measurement and model. The agreement between the measured and simulated EQE of the green device (55.8% versus 55.9%) supports the validity of the model inputs.

## OLED fabrication and characterization

The green OLED has the structure ITO (150 nm)/HATCN (5 nm)/TAPC (55 nm)/HATCN (5 nm)/TAPC (55 nm)/HATCN (5 nm)/TAPC (55 nm)/TCTA (10 nm)/TCTA:B3PyMPM:Ir(ppy)$_2$acac (25 nm, 4 wt%, co-evaporation)/B3PyMPM (50 nm)/B3PyMPM:$Cs_2CO_3$ (5 nm, 20 wt%)/Ag (100 nm) and was encapsulated with a thin-film nanolaminate of alternating $Al_2O_3$ and $ZrO_2$ deposited by atomic layer deposition. A 3-μm Parylene-C coating on top served both as an additional thin-film encapsulation and as the insulating layer that electrically separates the upper $MoO_x$/Ag reflector (5/100 nm) from the device below. A microlens-array film (MAF02, Lumtec) was used as the outcoupling structure, and index-matching oil (Immersion oil type-F, Olympus) removed the air gap to the glass substrate.

The orange OLED was fabricated on a high-index substrate patterned with a microlens array (HNPS01, $n$ = 1.77, Lumtec), on which 50 nm of IZO deposited by RF sputtering (123 W) served as the anode. On top of it, HATCN (5 nm)/TAPC (95 nm)/HATCN (5 nm)/TAPC (95 nm)/TCTA (10 nm)/TCTA:B3PyMPM:Ir(dmppy-ph)$_2$tmd (25 nm, 4 wt%, co-evaporation)/B3PyMPM (55 nm)/Liq (1 nm)/Al (100 nm) were deposited, and a DBR of 4.5 alternating ZnS (70 nm)/LiF (115 nm) pairs was deposited on the rear of the finished device. For comparison, an Al-coated Si wafer was placed behind a device with the same stack architecture using spacers to provide an Al auxiliary reflector (Supplementary Fig. 6). The ZnS and LiF thicknesses to be used were determined by particle swarm optimization (PSO) followed by a local gradient-based refinement, maximizing the simulated EQE of the complete device (Supplementary Note 2). Reference devices with the same stacks were fabricated on planar glass/IZO substrates.

All organic and metal layers and the ZnS/LiF DBR were deposited by thermal evaporation at a base pressure below $10^{-6}$ Torr, and all organic materials were purchased from Lumtec. The active areas were 4 $mm^2$ (green) and 9 $mm^2$ (orange), and all auxiliary reflectors measured 23 mm × 21 mm.

The devices were characterized in a custom-built goniometer set-up consisting of a source meter, a photodiode and a spectrometer. The EQE was obtained from the angle-resolved measurements by the method described previously[45,46], which accounts for any non-Lambertian emission of the device and thus avoids over- or underestimation of the EQE. The photodiode was placed 52 mm (green) and 59 mm (orange) from the device so that the entire substrate, including the microlens-array area outside the active area, lay within its field of view at all angles. Light emitted from the substrate edges was blocked, so that only light escaping from the front surface was counted. The current density and luminance were calculated using the active area of the device. The EQE is independent of the area, as it is obtained from the total number of photons emitted from the front surface, integrated over all angles, relative to the number of injected electrons.

## Acknowledgements

This work was supported by the Industry Technology R&D Program (RS-2025-25452683) funded by the Ministry of Trade, Industry and Energy (MOTIE, Korea), by the European Research Council under the European Union's Horizon Europe Framework Programme/ERC Advanced Grant agreement no. 101097878 (HyAngle, to M.C.G.) and by the Alexander von Humboldt Foundation (Humboldt Professorship to M.C.G.). J.K. acknowledges funding from the European Union's Horizon Europe research and innovation programme under the Marie Skłodowska-Curie grant agreement no. 101279265 (DBROLED).

## Competing interests

The authors declare no competing interests.

## Figures and Tables

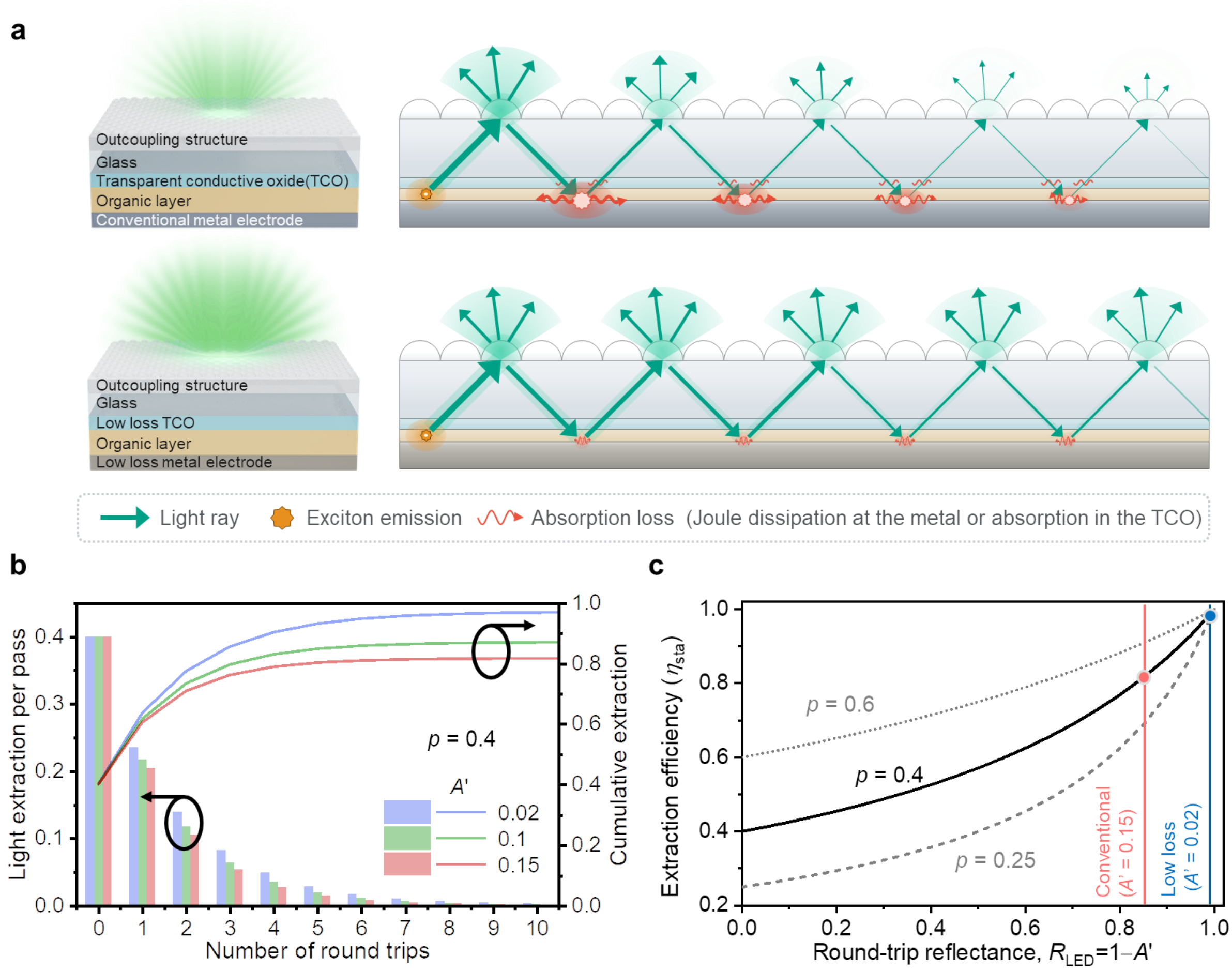

**Figure 1. Light extraction efficiency depends on the round-trip loss inside the OLED.** (a) Schematic of light extraction with an outcoupling structure attached to a conventional OLED (top) and to an OLED designed for minimal round-trip loss (bottom). The light delivered to the substrate ($P_{\mathrm{sub}}$) is partly transmitted by the outcoupling structure ($B_{\mathrm{T}}$) and partly returned ($B_{\mathrm{R}}$), reflected by the device stack ($R_{\mathrm{LED}}$) and sent back to the outcoupling structure. For every round trip, a fraction $A' = 1 - R_{\mathrm{LED}}$ is lost to Joule dissipation at the metal electrode and to absorption in the transparent conductive oxide (TCO) electrode. (b) Light extraction per pass (bars) and cumulative extracted fraction (lines) versus the number of round trips for an escape probability $p = 0.4$ and for round-trip losses $A' = 0.02, 0.10$ and 0.15. (c) Extraction efficiency versus the round-trip reflectance $R_{\mathrm{LED}} = 1 - A'$ for $p = 0.25, 0.4$ and 0.6 (assuming no transmission through the rear of the device). The red and blue vertical lines indicate $R_{\mathrm{LED}}$ for a conventional and a low-loss design, respectively.

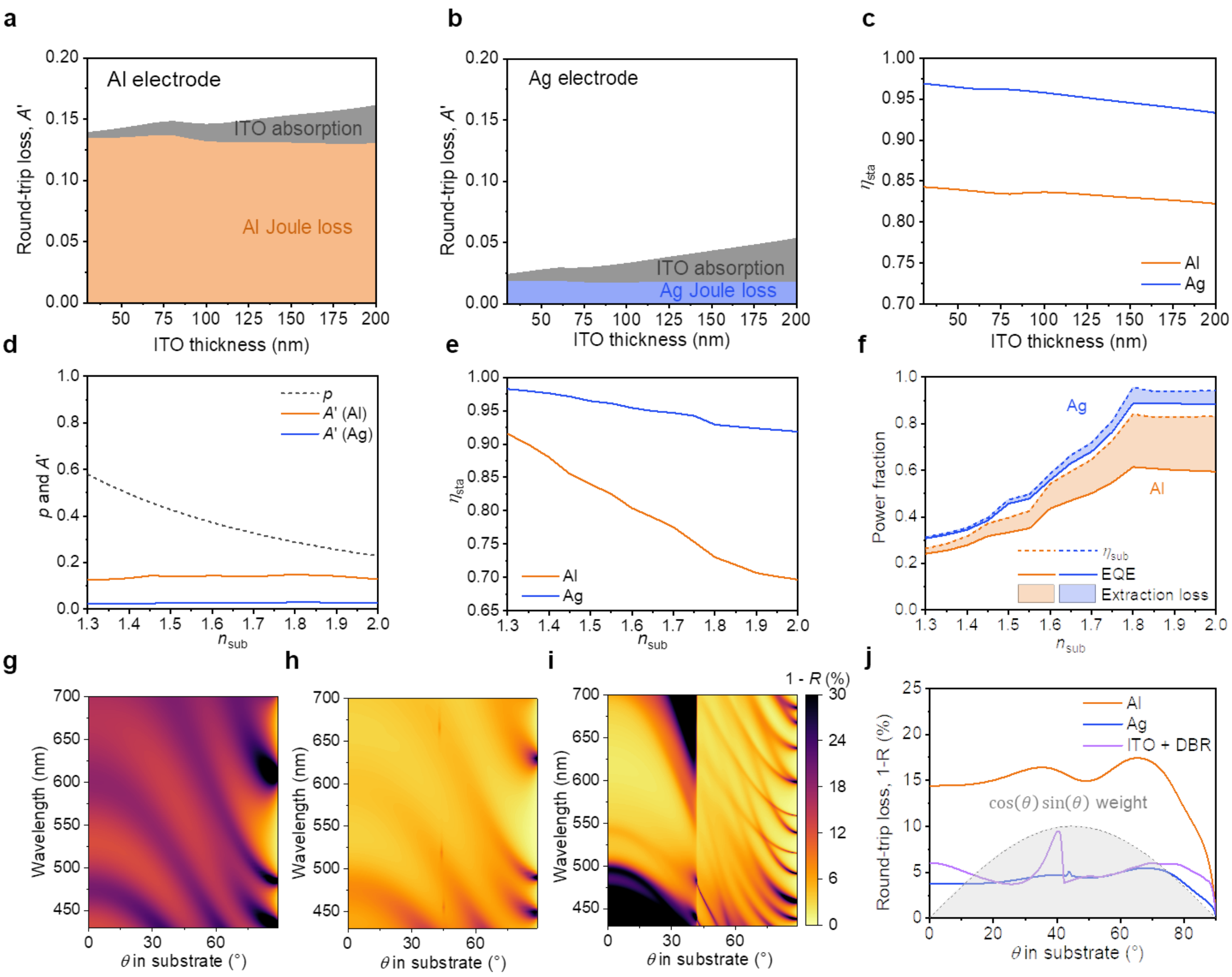

**Figure 2. Round-trip loss, substrate-delivery efficiency and EQE for different device structures.** (a, b) Round-trip loss $A'$ suffered by light returned to the OLED by an outcoupling structure, split into the Joule dissipation loss at the metal reflector (orange and blue) and the ITO absorption (gray) as a function of ITO thickness, for Al (a) and Ag (b) reflectors. (c) Resulting substrate-to-air extraction efficiency $\eta_{\mathrm{sta}}$. (d) Escape probability $p$ and round-trip loss $A'$, (e) substrate-to-air extraction efficiency $\eta_{\mathrm{sta}}$, (f) substrate-delivery efficiency $\eta_{\mathrm{sub}}$ (dashed), EQE (solid) and their difference, the extraction loss (shaded), all as a function of the substrate refractive index. (g–i) Angle- and wavelength-resolved round-trip loss A′ for a 150-nm-thick ITO transparent electrode and different reflectors: Al (g), Ag (h) and a 50-nm ITO cathode with a ten-pair ZnS/LiF DBR (i); (g–i) share the same false-color scale. (j) Round-trip loss versus angle averaged over the emission spectrum for Al (orange), Ag (blue) and ITO with ten-pair ZnS/LiF DBR (purple), with the angular weight $\cos\theta\sin\theta$ (shaded, a.u.). The $\eta_{\mathrm{sta}}$ and EQE in (c), (e) and (f) are obtained from equation (2) using values of $p$ and $A'$ from (a), (b) and (d); the $p$ in (d) is the single-pass escape probability obtained by averaging the transmitted component of the bidirectional scattering distribution function (BSDF) of a hemispherical microlens array with the same index as the substrate, using the angular weight $\cos\theta\sin\theta$ (0.43 at $n$ = 1.5, 0.29 at 1.8).

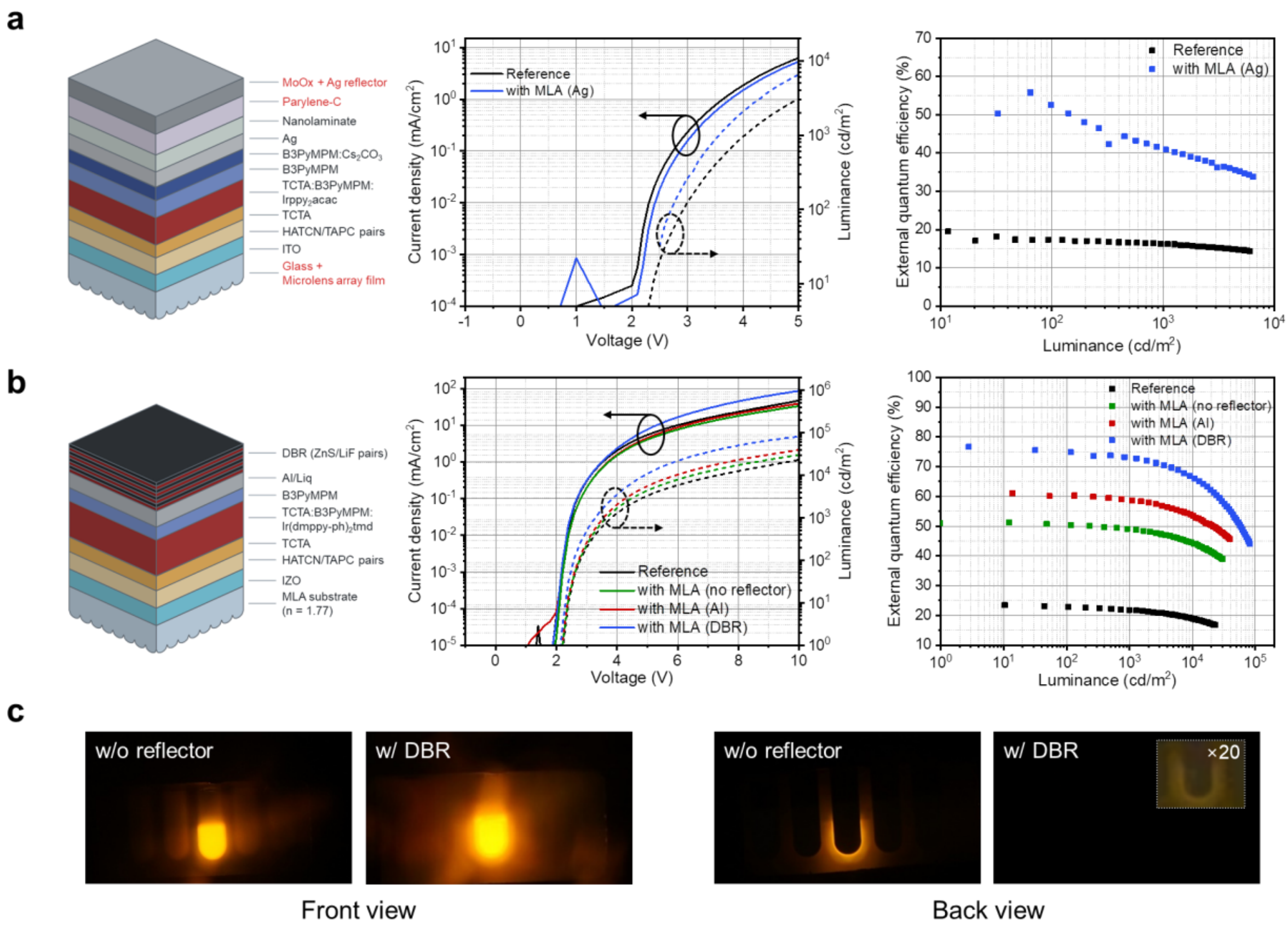

**Figure 3. Performance of OLEDs with microlens array and auxiliary reflector.** (a) Device structure, *J–V–L* and *L–EQE* characteristics of green-emitting phosphorescent OLEDs fabricated on a glass substrate with a microlens-array (MLA) film, with a 100-nm Ag cathode as low-loss reflector and an additional Ag film larger than the active area deposited as auxiliary reflector onto the thin-film encapsulation. Maximum EQE: 20% for the planar reference and 55.8% with the microlens-array film and the auxiliary Ag reflector. (b) Device structure, *J–V–L* and *L–EQE* characteristics of orange-emitting phosphorescent OLEDs fabricated on a high-index substrate ($n_{\mathrm{sub}}$ = 1.77) with a microlens array and either an Al or a ZnS/LiF DBR auxiliary reflector on the back. Maximum EQE: planar reference device 23%, microlens-array substrate without auxiliary reflector 51%, with Al auxiliary reflector 61% and with the DBR auxiliary reflector 77%. (c) Photographs of orange-emitting OLEDs without auxiliary reflector and with auxiliary DBR, viewed from the front and the back. The inset in the back-view image with the DBR shows the same image with the brightness increased 20-fold (×20).

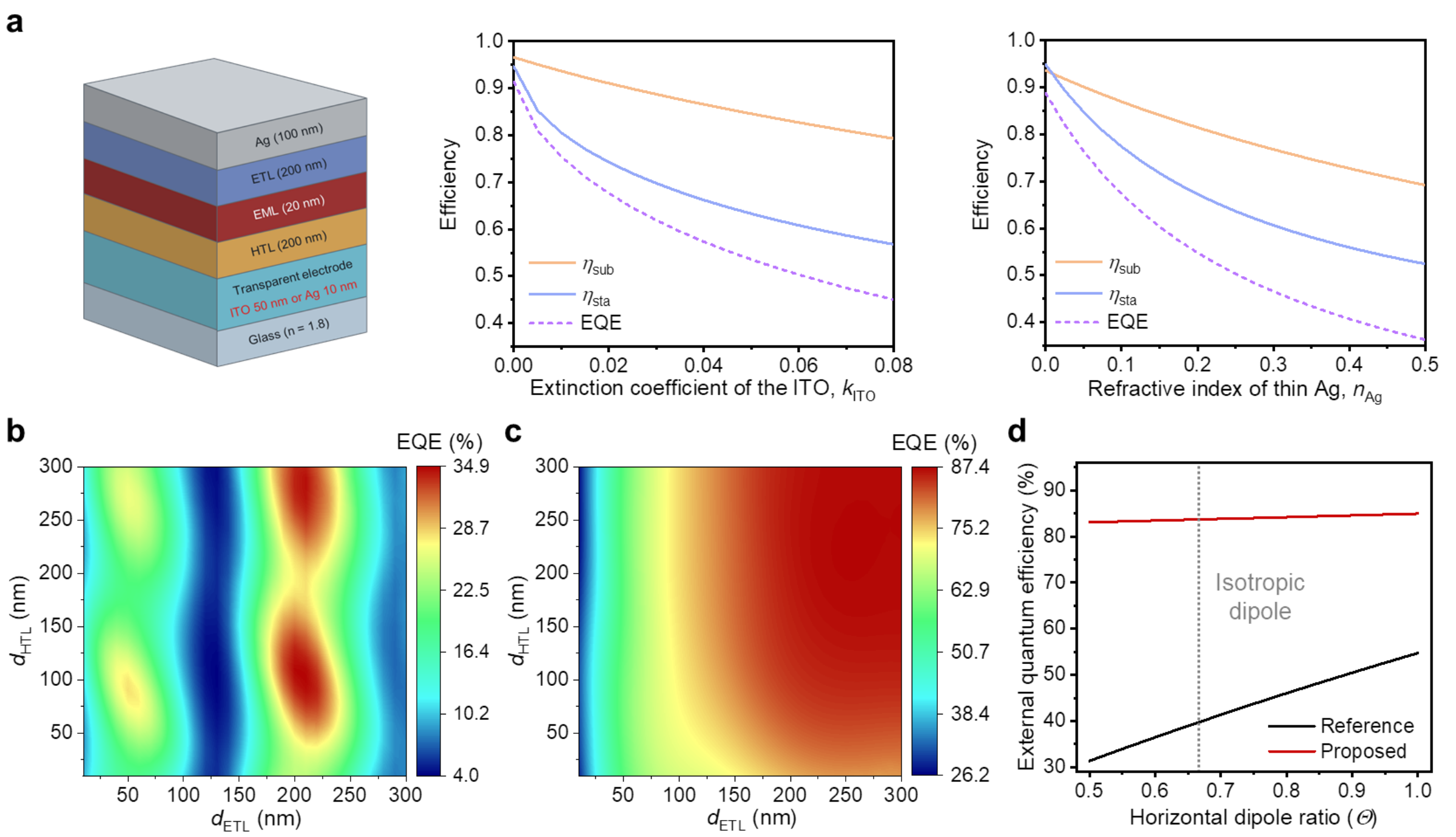

**Figure 4. Design freedom afforded by the proposed design rules.** (a) Substrate-delivery efficiency $\eta_{sub}$, substrate-to-air extraction efficiency $\eta_{sta}$ and EQE for the OLED structure on the left versus the main parameter affecting absorption for two different transparent electrodes, ITO (extinction coefficient, $k_{ITO}$, center) and 10 nm Ag (refractive index, $n_{Ag}$, right). For the calculation, the refractive index of ITO was fixed at $n_{ITO} =$ 1.86 (König et al.[40], 550 nm) and the extinction coefficient of Ag was fixed at $k_{Ag}$= 3.82. EQE versus electron-transport layer (ETL) and hole-transport layer (HTL) thickness for (b) a conventional OLED (Al cathode, $n_{sub}$ = 1.5, no outcoupling structure) and (c) a device following the proposed design rules (Ag reflector, $n_{sub}$ = 1.8, microlens array). (d) EQE versus the horizontal dipole ratio $\Theta$ of the emitter for a reference OLED with Al cathode and stack thickness tuned for optimal outcoupling (ETL/HTL 80/230 nm) on glass substrate ($n_{sub}$ = 1.5) with microlens array and for a device following the proposed design rules with Ag cathode and unoptimized thicknesses (ETL/HTL 200/200 nm) on a high-index microlens substrate ($n_{sub}$ = 1.8). The dotted vertical line indicates an isotropic emitter orientation ($\Theta$ = 2/3). (a), (c) and (d) were computed with the angle-resolved model (Methods) using the BSDF of a hemispherical microlens array with $n_{MLA}$ = 1.8. (a) and (d) were computed at a single wavelength of 550 nm assuming PLQY = 1 and (a) further used $\Theta$ = 2/3; (b) and (c) use the emission spectrum of $Ir(ppy)_2acac$ and assume PLQY = 1 and $\Theta$ = 2/3.

*Supplementary Information*

# Minimizing round-trip loss for near-unity quantum efficiency in organic light-emitting diodes

Junho Kim[1,2], Seonjeong Lee[3], Seunghyup Yoo[2,3], Malte C. Gather[1,4]

[1] Humboldt Centre for Nano- and Biophotonics, Department of Chemistry and Biochemistry, University of Cologne, Köln, Germany.

[2] School of Electrical Engineering, Korea Advanced Institute of Science and Technology (KAIST), Daejeon, Republic of Korea.

[3] Graduate School of Semiconductor Technology, Korea Advanced Institute of Science and Technology (KAIST), Daejeon, Republic of Korea.

[4] School of Physics and Astronomy, University of St Andrews, North Haugh, St Andrews, UK.

Correspondence: Seunghyup Yoo (syoo_ee@kaist.ac.kr), Malte C. Gather (malte.gather@uni-koeln.de)

**Supplementary Note 1. Multi-pass extraction model**

An OLED with an external outcoupling structure consists of a coherent thin-film stack (a few hundred nanometers to about a micrometer thick) and an incoherent substrate and outcoupling structure (hundreds of micrometers to millimeters thick). We treat the thin-film stack with wave optics, i.e., the classical dipole model[1,2] and the transfer-matrix method, and the substrate and outcoupling structure with ray tracing. The two regions are coupled only through the power propagating in the substrate[3,4] at an angle $\theta$. For a semi-infinite substrate, the wave-optical calculation gives the fraction of the power generated in the emissive layer that is delivered to the substrate, $\eta_{\mathrm{sub}}(\lambda)$, its angular distribution, $P_{\mathrm{sub}}(\theta,\lambda)$, normalized such that $\sum_\theta P_{\mathrm{sub}} = 1$, and the polarization-averaged reflectance of the device for light incident from the substrate, $R_{\mathrm{LED}}(\theta,\lambda)$. Ray tracing gives the bidirectional scattering distribution function (BSDF) of the outcoupling structure. The BSDF generally depends on the incident direction ($\theta_{\mathrm{i}}$, $\varphi_{\mathrm{i}}$), the scattered direction ($\theta_{\mathrm{s}}$, $\varphi_{\mathrm{s}}$) and the wavelength. For microlens arrays and isotropic scattering layers, however, it can be taken as azimuthally symmetric and reduces to a function of the incident and scattered polar angles. In addition, since the refractive indices of common substrates and outcoupling layers vary little across the visible range, we used a BSDF calculated at 550 nm for all visible wavelengths:

$$BSDF(\theta_i, \varphi_i, \theta_s, \varphi_s, \lambda) \rightarrow B(\theta_i, \theta_s) \quad \text{(S1)}$$

The BSDF is divided into a transmitted component (the bidirectional transmittance distribution function, BTDF, for scattering angles of 0–90°) and a reflected component (the bidirectional reflectance distribution function, BRDF, for 90–180°). From these, we define the fraction transmitted to air at each angle of incidence and the angle-resolved fraction returned to the substrate, $B_{\mathrm{T}}(\theta_i)$ and $B_{\mathrm{R}}(\theta_i, \theta_r)$. All angular quantities are discretized into polar-angle bins, so that the BTDF and BRDF give the fraction of the incident power scattered into each bin. In equations (S2)–(S6), the transmitted light is indexed by its exit angle in air and the returned light by its propagation angle in the substrate (i.e., 180° minus the reflected angle plotted in Supplementary Fig. 1):

$$B_{\mathrm{T}}(\theta_i) = \sum_{\theta_t} BTDF(\theta_i, \theta_t) \quad \text{(S2)}$$

$$B_{\mathrm{R}}(\theta_i, \theta_r) = BRDF(\theta_i, \theta_r) \quad \text{(S3)}$$

For a non-absorbing outcoupling structure, the two satisfy the following relation:

$$\sum_{\theta_r} B_{\mathrm{R}}(\theta_i, \theta_r) = 1 - B_{\mathrm{T}}(\theta_i) \quad \text{(S4)}$$

Each time the light in the substrate encounters the outcoupling structure, a fraction of it is extracted. The remainder is redirected back to the device, reflected with reflectance $R_{\mathrm{LED}}$ and sent to the outcoupling structure again. Denoting the angular distribution of the light that first reaches the outcoupling structure as $P_{\mathrm{sub}}^{(0)}(\theta)$, we obtain the distribution at each subsequent pass from the recursion below:

$$P_{\mathrm{sub}}^{(0)}(\theta) = P_{\mathrm{sub}}(\theta), \qquad \sum_{\theta} P_{\mathrm{sub}}(\theta) = 1 \tag{S5}$$

$$P_{\mathrm{sub}}^{(k+1)}(\theta_r) = R_{\mathrm{LED}}(\theta_r) \sum_{\theta_i} P_{\mathrm{sub}}^{(k)}(\theta_i) B_{\mathrm{R}}(\theta_i, \theta_r) \tag{S6}$$

The power extracted at the $k$-th pass, $\eta_{\mathrm{sta}}^{(k)}$, and the extraction efficiency summed over all passes, $\eta_{\mathrm{sta}}$, are then given by

$$\eta_{\mathrm{sta}}^{(k)} = \sum_{\theta} P_{\mathrm{sub}}^{(k)}(\theta) B_{\mathrm{T}}(\theta) \tag{S7}$$

$$\eta_{\mathrm{sta}} = \sum_{k=0}^{\infty} \eta_{\mathrm{sta}}^{(k)} = P_{\mathrm{sub}}(I - B_{\mathrm{R}} R)^{-1} {B_{\mathrm{T}}}^{\mathrm{T}} \tag{S8}$$

where $P_{\mathrm{sub}}$ and $B_{\mathrm{T}}$ are row vectors, $B_{\mathrm{R}}$ is a matrix and $R = \mathrm{diag}[R_{\mathrm{LED}}(\theta)]$. Because part of the light is transmitted to air at every pass and the device reflectance is below unity, the spectral radius of $B_{\mathrm{R}} R$ is smaller than unity and the series converges. The fraction of photons generated in the emissive layer that escape to air is then obtained by weighting with $w(\lambda)$, the emission spectrum normalized to photon number. At a single wavelength, this reduces to $\eta_{\mathrm{out}} = \eta_{\mathrm{sub}} \cdot \eta_{\mathrm{sta}}$.

$$\eta_{\mathrm{out}} = \sum_{\lambda} w(\lambda) \eta_{\mathrm{sub}}(\lambda) \eta_{\mathrm{sta}}(\lambda) \tag{S9}$$

With unity charge balance and exciton utilization and with the effective PLQY included in the calculation of $\eta_{\mathrm{sub}}$, this gives equation (1) of the main text. When the full emission spectrum is used, the substrate-delivery efficiency in equation (1) is its average weighted by the photon-number PL spectrum as in equation (S9), and the extraction efficiency is the ratio of equation (S9) to this average. The first factor, $\eta_{\mathrm{sub}}$, thus reflects only how the power is shared among the modes inside the device (waveguided modes, surface plasmon polaritons and absorption), and $\eta_{\mathrm{sta}}$ only how the light delivered to the substrate is consumed between the device and the outcoupling structure.

If the angle of the returned light is assumed to be randomized at every pass, the angular distribution of the light reaching the outcoupling structure becomes a single distribution $g(\theta)$, normalized such that its sum over all angles is unity, that is independent of the pass number, and all angle-dependent quantities can be replaced by their averages over this distribution.

$$p = \sum_{\theta} g(\theta) B_{\mathrm{T}}(\theta) \tag{S10}$$

$$A' = 1 - \sum_{\theta} g(\theta) R_{\mathrm{LED}}(\theta) \tag{S11}$$

For a non-absorbing outcoupling structure, the returned fraction is $1 - p$. The power extracted at the $k$-th pass is then $p[(1-p)(1-A')]^k$, and its sum gives equation (S12), i.e., equation (2) of the main text.

$$\eta_{\mathrm{sta}} = p \sum_{k \geq 0} [(1-p)(1-A')]^k = \frac{p}{[p + (1-p)A']} \tag{S12}$$

In the main text, we use $g(\theta) \propto \cos\theta \sin\theta$, i.e., the weighting of a Lambertian distribution in the substrate. When the reflector partially transmits light (for example, a dielectric reflector), the rear transmission is also counted as light that does not return, so that $A' = 1 - R_{\mathrm{LED}}$ includes both absorption and transmission.

Equation (S12) rests on two assumptions. First, the light returned by the outcoupling structure is randomized in angle at every pass, with the distribution $\cos\theta \sin\theta$. Second, the distribution of the light that the device first delivers to the substrate, $P_{\mathrm{sub}}(\theta)$, follows the same form. Because of these assumptions, equation (S12) differs from the angle-resolved model of equation (S8), as discussed below. As shown in Supplementary Figs 1 and 2, the angle-resolved transmittance of a widely used hexagonally close-packed hemispherical microlens array is high below about 60°, peaks at about 50–57° and falls below 0.1 at angles above about 72°. In addition, as the BRDF shows (Supplementary Fig. 1), light incident at large angles tends to be returned at large angles. Because these BSDF characteristics do not change from pass to pass, light at favorable angles below about 60° is extracted mostly in the first passes, and the remaining light escapes less efficiently. The distribution $P_{\mathrm{sub}}^{(k)}$ therefore evolves. The evolved distribution consists of light at unfavorable angles, and $p$ decreases progressively. As a result, $\eta_{\mathrm{sta}}$ obtained from the series is generally lower than that given by equation (2). The effective escape probability of a real device can therefore be approximated by $\bar{p}$, the average of $p_k$ weighted by the power reaching the outcoupling structure at each pass, which is generally lower than $p$ in equation (2), evaluated for a Lambertian distribution (Supplementary Fig. 3a,b). Depending on the device structure, however, the difference between the two can become small when the light delivered to the substrate is concentrated at angles that escape easily, or when the outcoupling structure, such as a scattering layer, sufficiently mixes the returned light. Because equation (2) captures the trends with only two quantities, $p$ and $A'$, independently of the detailed angular characteristics of the outcoupling structure and the device, it was used in the main text to illustrate design trends (Figs 1 and 2c–f of the main text). The quantitative values for specific devices (Figs 4a,c,d of the main text, Supplementary Figs 11 and 12 and Supplementary Table 1) were calculated with the angle-resolved model.

The deviation also depends on the substrate index (Supplementary Fig. 3c). At 550 nm, the angle-resolved model falls below equation (2) by up to about 9 (Ag) and 12 (Al) percentage points just above each substrate index at which a waveguided mode enters the substrate at a near-grazing angle (about 1.42–1.46, 1.58–1.59 and 1.75–1.78, depending on the cathode, and at 1.8, where the substrate index matches that of the organic layers). At these indices, a large part of the light delivered to the substrate propagates at grazing angles, where the microlens array transmits little (Supplementary Fig. 2), and its angular distribution deviates strongly from a Lambertian one. These dips do not reduce the EQE, because this light raises the substrate-delivery efficiency (Fig. 2f of the main text). The EQE obtained with the series therefore increases almost monotonically with the substrate index (Supplementary Fig. 3d). Between these indices, the difference is about 2–6 percentage points. Below an index of about 1.35, the two agree within 1 percentage point, and above about 1.9 the series is slightly higher than equation (2), by up to about 2 percentage points. The positions of the dips depend on the mode structure of the specific stack. In the real material stacks of Supplementary Table 1, which contain birefringent layers with lower extraordinary indices and different thicknesses, a substrate index of 1.8 does not lie just above a cut-off, and equation (2) exceeds the series by only 1.6–3.2 percentage points in both the extraction efficiency and the EQE.

**Supplementary Note 2. Angle- and spectrum-robust DBR design**

A conventional dielectric DBR consists of alternating high- and low-index layers, each with an optical thickness of a quarter of the design wavelength. This quarter-wave condition holds only at normal incidence. At an oblique angle $\theta$, the phase in each layer depends on the propagation angle $\theta_\mathrm{j}$ in that layer, and high reflectance at the same wavelength would require a layer thickness of $d_\mathrm{j} = \lambda/(4n_\mathrm{j}\cos\theta_\mathrm{j})$. The high-reflectance band therefore shifts to shorter wavelengths with increasing angle of incidence. In an OLED with an external outcoupling structure, however, the reflector has to reflect all of the light making round trips in the substrate, not only normally incident light, i.e., it must be highly reflective over the full angle- and wavelength-resolved power distribution. For the calculations of Fig. 2i,j of the main text, we therefore designed the DBR as a chirped stack, in which the layer thicknesses vary gradually, and optimized it for the maximum reflectance averaged over the emission spectrum and the angular distribution.

Supplementary Fig. 5 compares a DBR designed with the quarter-wave condition at 550 nm with a chirped DBR optimized for the emission spectrum of an $Ir(ppy)_2acac$-based OLED and a Lambertian angular distribution of the light delivered to the substrate (10 ZnS/LiF pairs directly on a substrate with $n$ = 1.5, without ITO or organic layers, and with air behind the stack). At normal incidence, the quarter-wave DBR reflects more than 90% over a broad band from 474 to 654 nm, but as the angle of incidence increases its high-reflectance band shifts to shorter wavelengths, narrowing to 422–506 nm at 40°. The band in which the reflectance stays above 90% at all angles that can escape to air (0–41.8°) therefore spans only 474–500 nm (about 26 nm), much narrower than the PL spectrum of a typical emitter. Its reflectance averaged over the Lambertian angular distribution and the emission spectrum is 93.3%, and up to 39% of the light at 30–41° is transmitted (averaged over the spectrum). In the chirped DBR, by contrast, the progressively thicker layers reflect different wavelength ranges, and the high-reflectance band becomes broader. While its normal-incidence reflectance remains comparable to that of the quarter-wave DBR (97.3% for the chirped DBR versus 99.4% for the quarter-wave DBR, averaged over the spectrum), the band with more than 90% reflectance at all escape angles widens to 508–596 nm, about 88 nm. This band covers most of the typical emission band of a single-color OLED, and the average reflectance increases to 98.3%. The optimized thicknesses increase linearly along the stack, from 57 to 80 nm for ZnS and from 88 to 124 nm for LiF. These values refer to the DBR alone. In the device configuration of Fig. 2i,j of the main text, where the same DBR is placed behind a 50-nm ITO cathode, the organic layers and a 150-nm ITO anode, the rear transmission is 1.0% and the two ITO layers absorb 4.1%, giving a round-trip loss of 5.1%. The DBR of the orange device in Fig. 3b of the main text is a simpler uniform stack of 4.5 ZnS (70 nm)/LiF (115 nm) pairs, whose thicknesses were optimized for the EQE of this device by particle swarm optimization (PSO), as described below.

**Optimization of the 4.5-pair DBR used in the experiment.** The DBR deposited on the orange device (Fig. 3b of the main text) is a uniform stack in which all ZnS layers and all LiF layers have the same thickness. The two thicknesses were optimized by particle swarm optimization (PSO)[5,6]. The objective function was the simulated EQE of the complete orange device with the DBR on its rear, calculated with the angle-resolved model (Methods of the main text) from the measured optical constants, the microlens-array BSDF and the photoluminescence spectrum of $Ir(dmppy\text{-}ph)_2tmd$; the DBR reflectance enters this calculation through the device reflectance computed with the transfer-matrix method for $s$ and $p$ polarization. Each particle encoded the pair ($d_{\mathrm{ZnS}}$, $d_{\mathrm{LiF}}$) within 10–200 nm. The optimization used the particleswarm function of the MATLAB Global Optimization Toolbox with a swarm of 300 particles and a maximum of 20 iterations; all other settings were the default values (initial positions drawn uniformly within the bounds, adaptive inertia weight within 0.1–1.1, self- and social-adjustment weights of 1.49, a minimum neighborhood fraction of 0.25, and termination when the best objective changed by less than $10^{-6}$ over 20 consecutive iterations). The best particle was then refined by a local gradient-based search (fmincon, MATLAB Optimization Toolbox) as a hybrid function. The optimum was ZnS 70 nm and LiF 115 nm. Unlike the chirped design of Supplementary Fig. 5, which was optimized for the emission spectrum of $Ir(ppy)_2acac$, this DBR is tailored to the orange emitter of the experimental device.

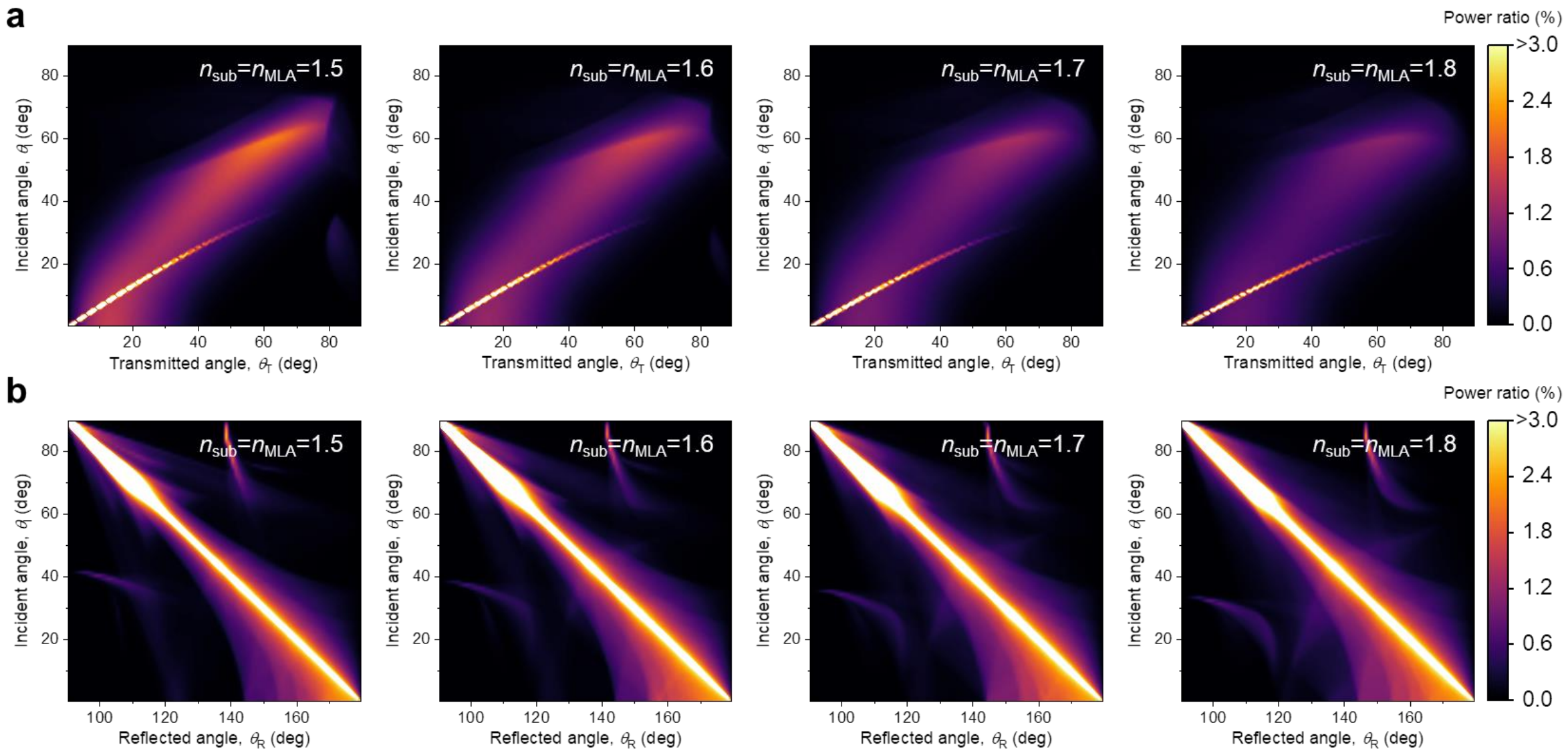

**Supplementary Fig. 1** Bidirectional scattering distribution function of a hexagonally close-packed hemispherical microlens array for different refractive indices. (a) BTDF and (b) BRDF for $n_{\mathrm{sub}}$ = $n_{\mathrm{MLA}}$ = 1.5, 1.6, 1.7 and 1.8. Light is incident on the microlens array from a substrate with the same refractive index. The transmitted light is plotted against the exit angle in air and the reflected light against the angle at which it re-enters the substrate. Each incident angle is normalized to the same incident power. The BSDFs for the other refractive indices used in the main text and in Supplementary Fig. 3 (e.g., 1.3, 1.4, 1.77, 1.9 and 2.0) were calculated in the same way.

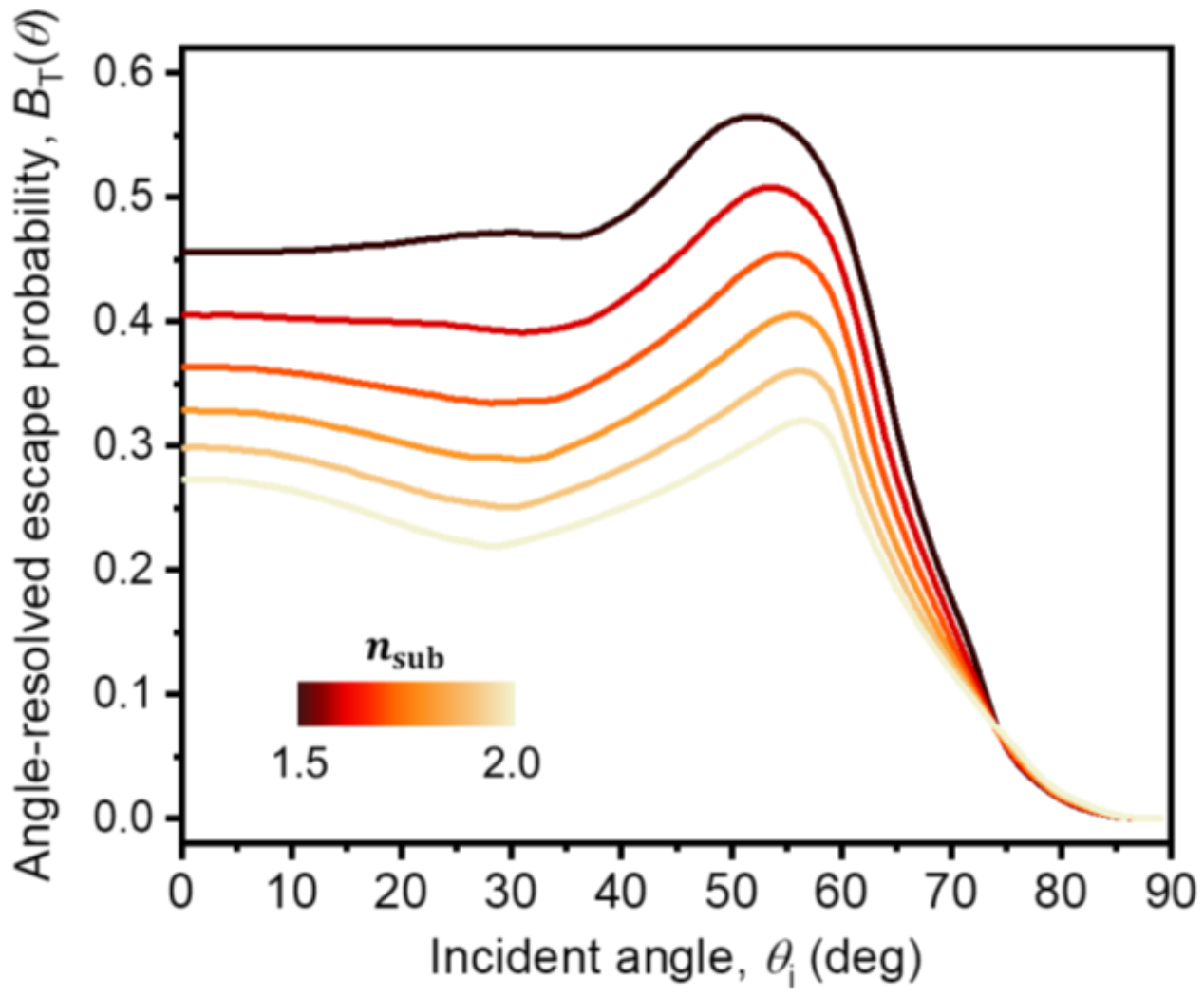

**Supplementary Fig. 2** Angle-resolved escape probability $B_T(\theta_i)$ of a hexagonally close-packed hemispherical microlens array for refractive indices $n_{sub}$ = $n_{MLA}$ = 1.5–2.0. $B_T(\theta_i)$ is the fraction of light incident on the microlens array at $\theta_i$ in the substrate that is transmitted to air in a single encounter, i.e., the BTDF of Supplementary Fig. 1 integrated over all exit angles. The escape probabilities for refractive indices of 1.3 and 1.4, used in Fig. 2d–f of the main text, were calculated in the same way.

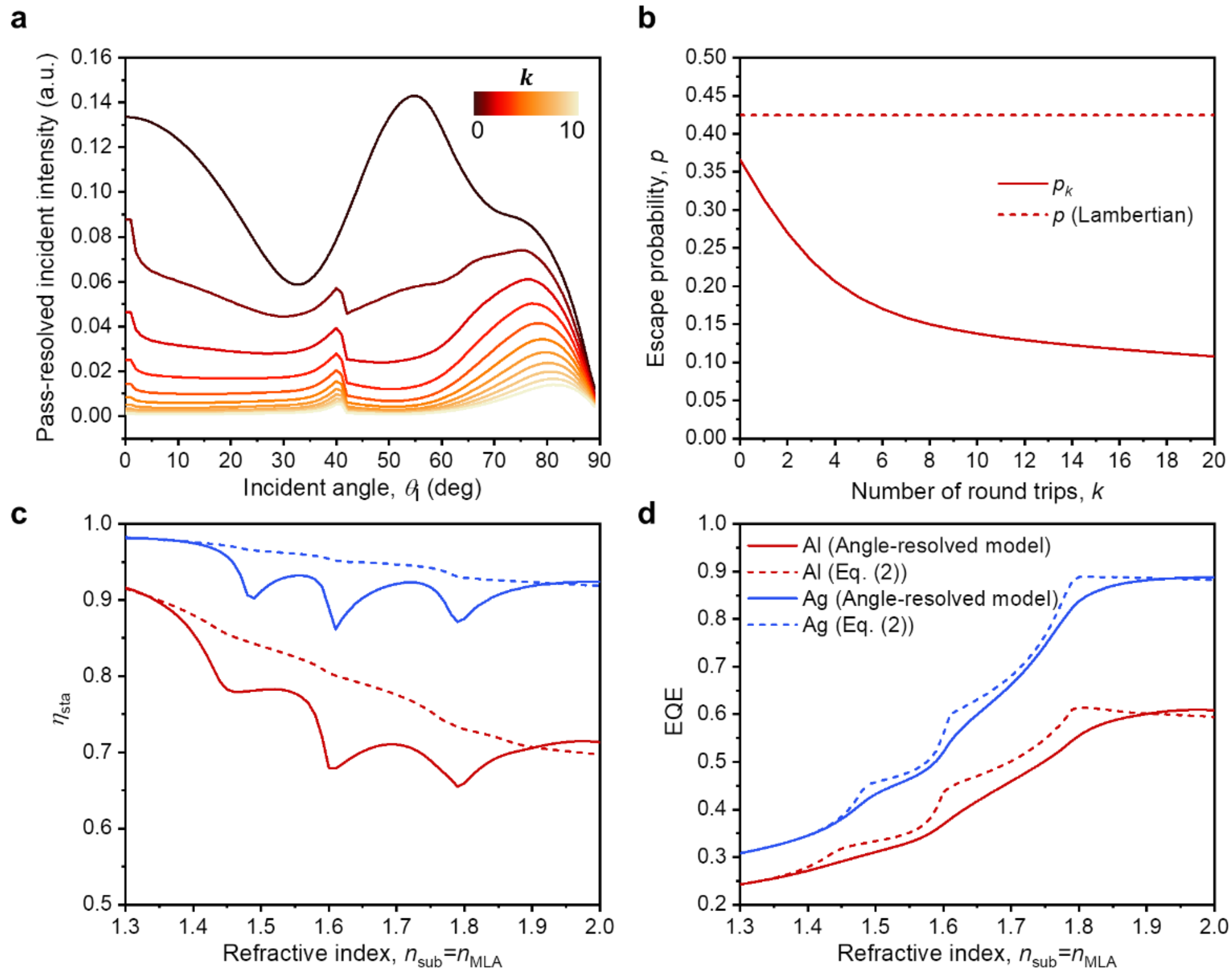

**Supplementary Fig. 3** Evolution of the light reaching the outcoupling structure with the number of round trips, and comparison of the angle-resolved model with equation (2). (a) Pass-resolved incident intensity (per unit solid angle) in the substrate of the light reaching the microlens array after $k$ round trips ($k$ = 0–10); $k$ = 0 is the substrate-delivered intensity of the device. (b) Escape probability $p_k$ of the light reaching the microlens array after $k$ round trips (solid line). The dashed line is the escape probability for a Lambertian distribution, $p$, used in equation (2). (a, b) Device: Ag (100 nm) / B3PyMPM (300 nm) / TCTA:B3PyMPM (25 nm) / TAPC (300 nm) / ITO (50 nm) / glass ($n$ = 1.5) with a hemispherical microlens array ($n$ = 1.5); isotropic emitter; Ir(ppy)$_2$acac emission spectrum (400–700 nm). (c) Substrate-to-air extraction efficiency and (d) EQE versus the refractive index of the substrate and the index-matched microlens array for Al and Ag reflectors, calculated with the angle-resolved model (solid lines) and with equation (2) (dashed lines). Stack as in Fig. 2d–f of the main text: substrate / ITO (50 nm) / organic layers (420 nm, $n$ = 1.8) / reflector (100 nm); 550 nm, isotropic emitter, PLQY = 1. The dips in (c) appear just above each index at which a waveguided mode enters the substrate, because the light from this mode propagates at near-grazing angles, where the microlens array transmits little. This light also raises the substrate-delivery efficiency (Fig. 2f of the main text), and the EQE in (d) therefore shows no corresponding dips. The positions of the dips depend on the mode structure of the specific stack (for the real material stack of Supplementary Table 1, the difference is 1.6–3.2 percentage points).

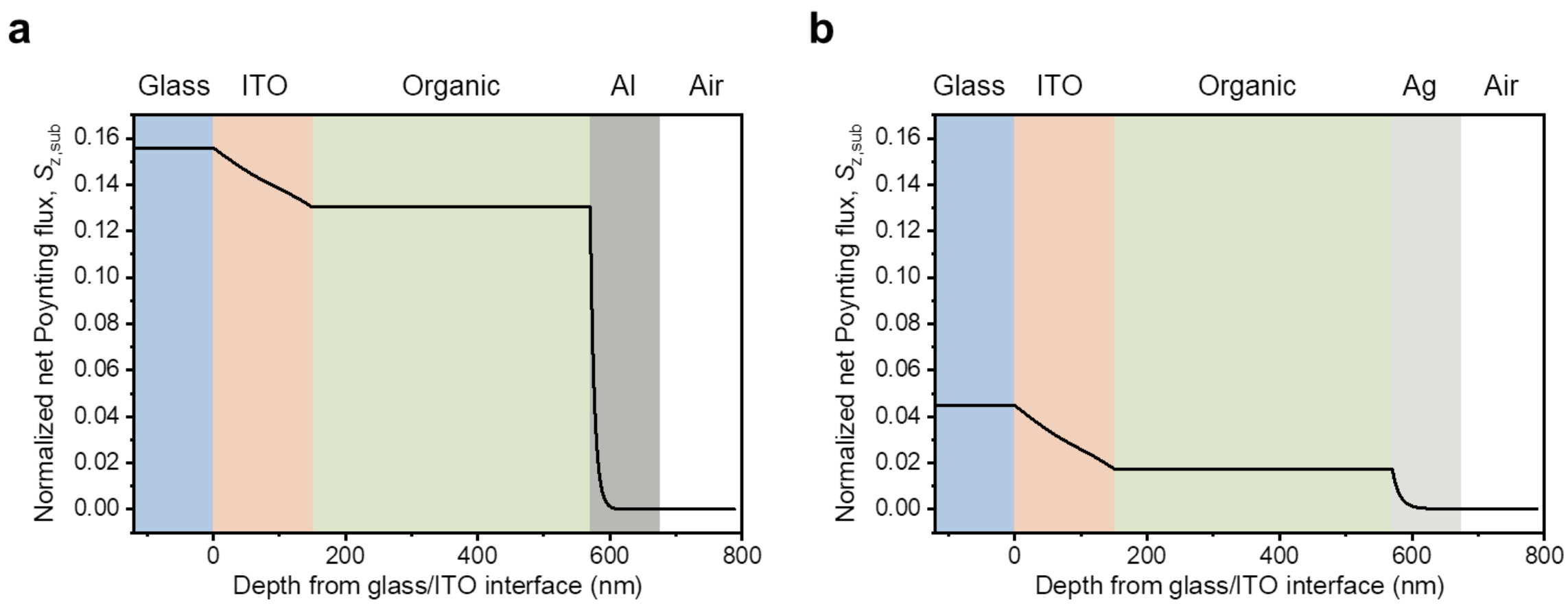

**Supplementary Fig. 4** Depth profile of the normalized net Poynting flux in OLEDs with (a) an Al and (b) an Ag reflector, for light returned from the substrate to the device. The flux is normalized to the power incident on the device from the substrate. Its value in the glass therefore equals the round-trip loss $A' = 1 - R_{\mathrm{LED}}$ (15.6% for Al, 4.5% for Ag), and each drop marks the absorption in that layer. Stack: glass ($n$ = 1.5) / ITO (150 nm) / organic layers (420 nm, $n$ = 1.8) / Al or Ag (100 nm) / air. The profiles are averaged over the Ir(ppy)$_2$acac emission spectrum (400–700 nm, photon-number weighting), over a Lambertian angular distribution in the substrate and over *s* and *p* polarizations. Optical constants: ITO from König et al.[7] and Ag from McPeak et al.[8].

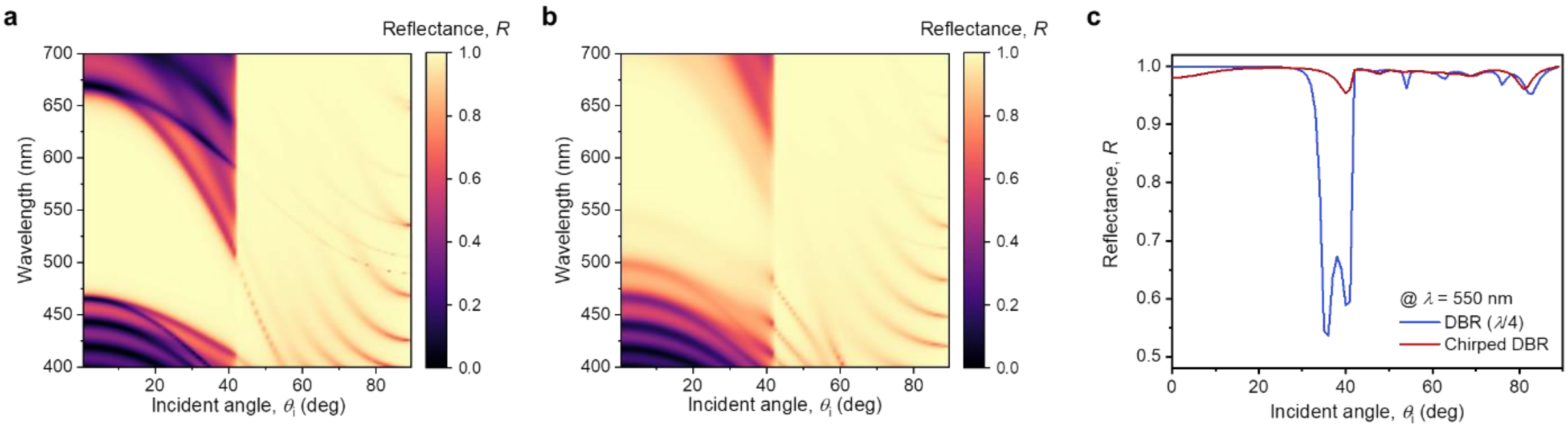

**Supplementary Fig. 5** Angle- and wavelength-resolved reflectance of (a) a quarter-wave DBR designed at 550 nm and (b) the optimized chirped DBR, for light incident from the substrate. (c) Reflectance of the two DBRs at 550 nm as a function of the incident angle. Both DBRs consist of ten ZnS/LiF pairs directly on a glass substrate ($n_{\mathrm{sub}}$ = 1.5), without ITO, with air behind the stack. Layer thicknesses: ZnS 58.2 nm / LiF 97.6 nm for the quarter-wave DBR; ZnS 57 → 80 nm / LiF 88 → 124 nm, increasing linearly along the stack, for the chirped DBR. Above the critical angle (41.8°), light is totally internally reflected at the rear surface.

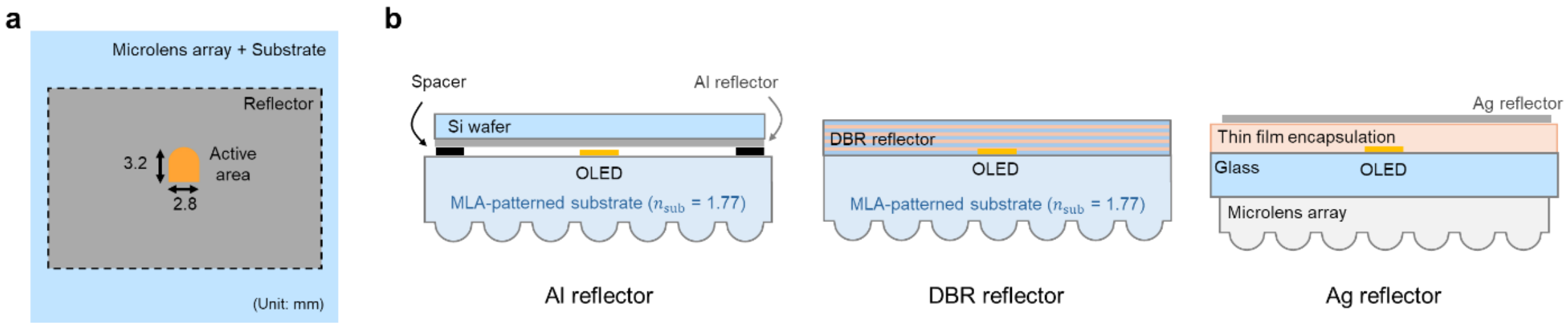

**Supplementary Fig. 6** Configurations used for the different auxiliary reflectors. (a) Top view of the device, showing the active area (4 mm$^2$ for the green and 9 mm$^2$ for the orange OLED) relative to the auxiliary reflector (23 mm × 21 mm) on a substrate. (b) Side views of the three auxiliary-reflector configurations: an Al-coated Si wafer placed behind the orange OLED on spacers (left), a 4.5-pair ZnS/LiF DBR deposited on the rear of the finished orange OLED (middle) and an Ag film deposited on the Parylene-C/thin-film encapsulation of the green OLED (right). Light returned by the microlens array spreads laterally in the thick substrate; the auxiliary reflectors, much wider than the active area, return the part that reaches the rear outside the electrode.

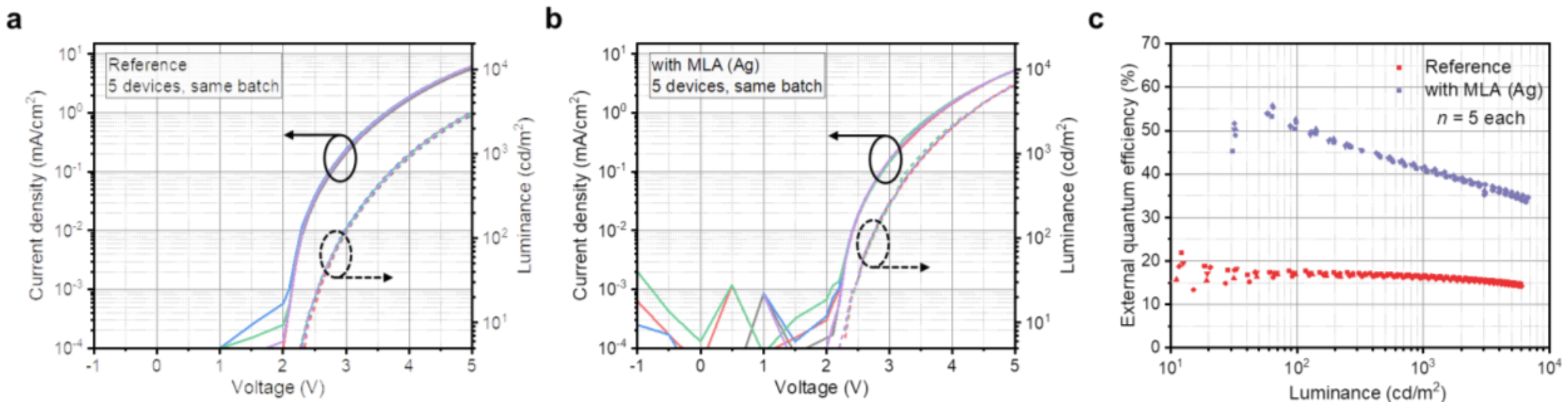

**Supplementary Fig. 7** Reproducibility summary for the green OLED structure shown in Fig. 3a of the main text. *J*–*V*–*L* characteristics of five reference devices on planar glass (a) and five devices with the microlens-array film and the Ag auxiliary reflector (b), fabricated in the same batch, and (c) *L*-EQE characteristics for the five devices from each group.

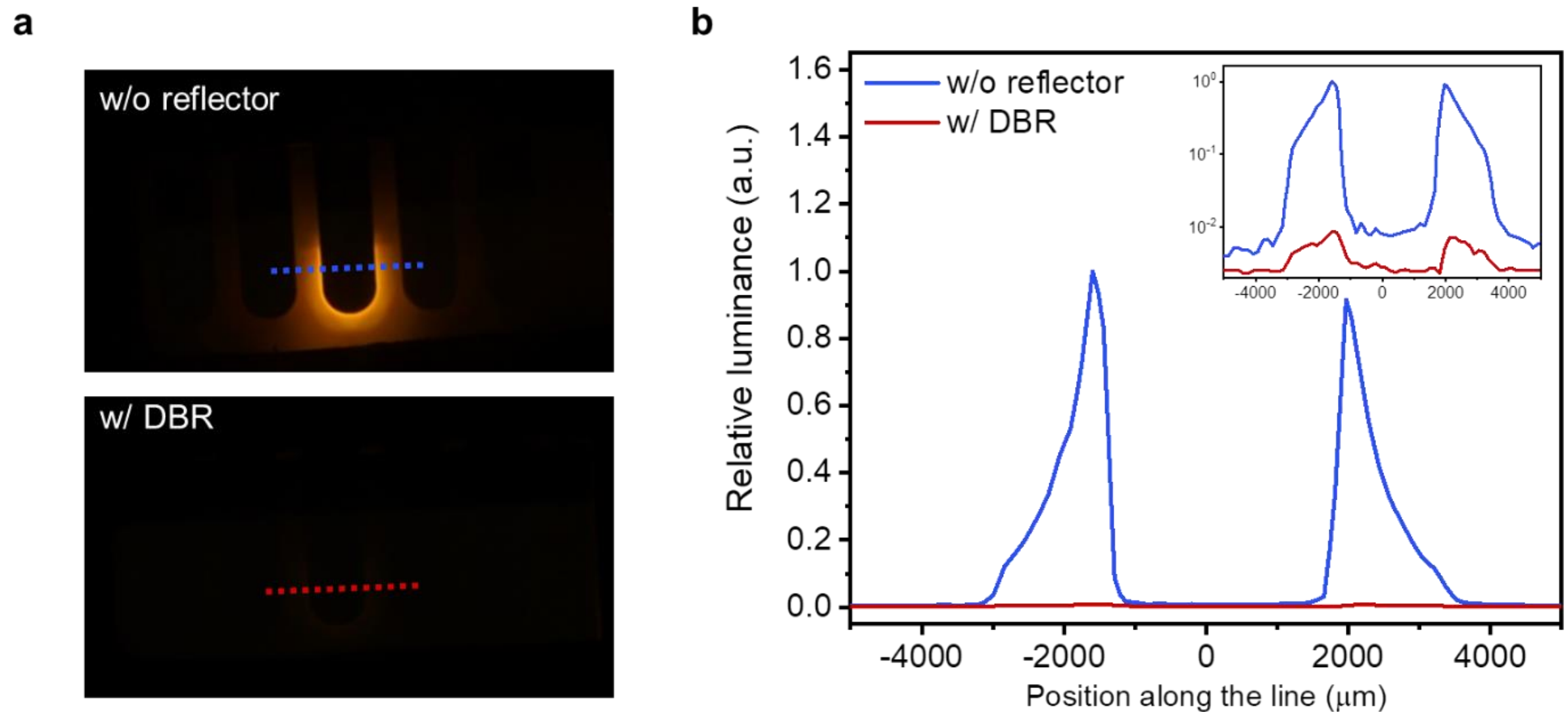

**Supplementary Fig. 8** Backside light leakage of the orange OLED without auxiliary reflector and with the DBR (Fig. 3c of the main text). (a) Back-view photographs taken under identical camera settings. (b) Relative luminance along the dotted lines in (a), normalized to the peak of the device without the reflector; inset, the same data on a logarithmic scale.

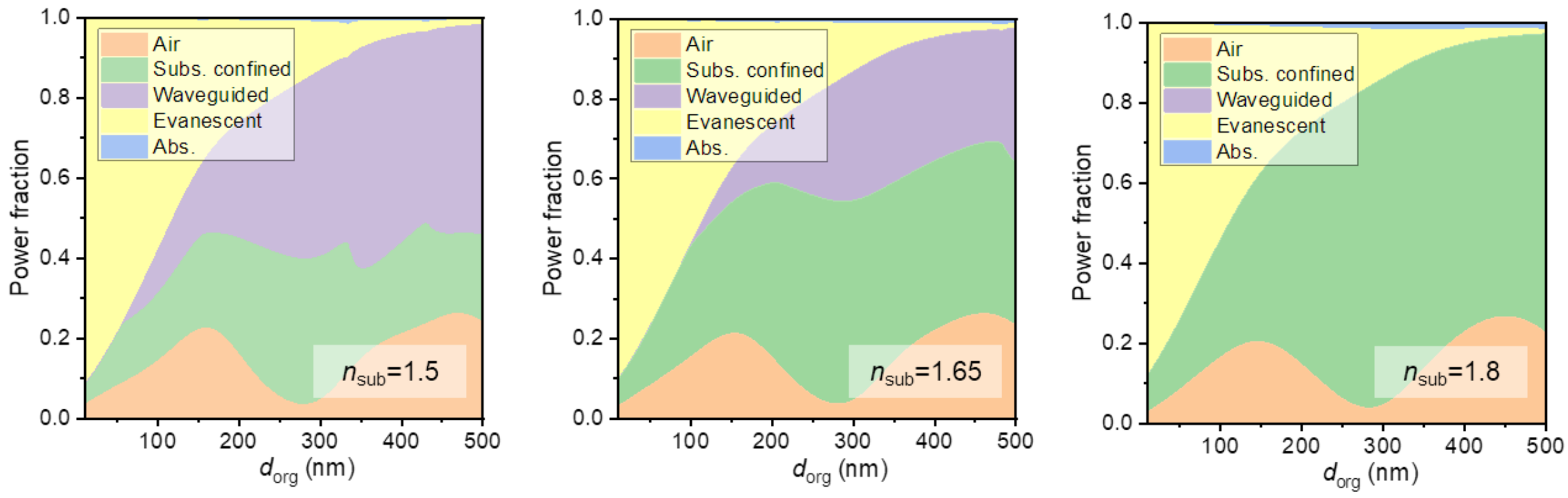

**Supplementary Fig. 9** Power fraction of each mode versus the total organic thickness $d_{org}$ (emitting layer at the center) for substrate indices $n_{sub}$ = 1.5, 1.65 and 1.8; Ag reflector, ITO 50 nm, organic layers with $n$ = 1.8.

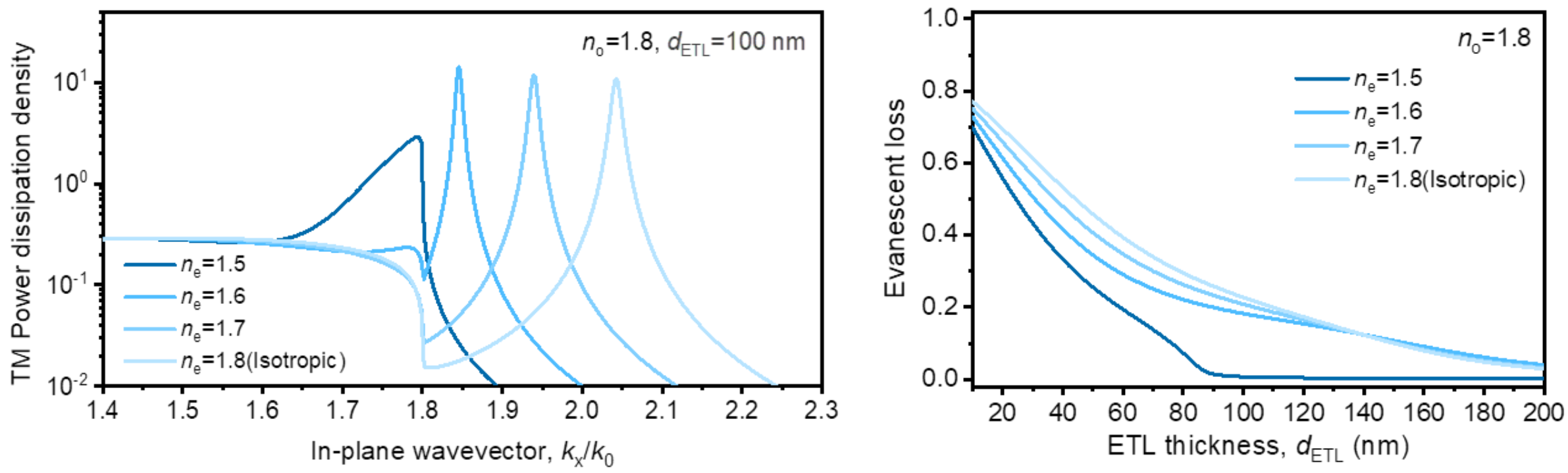

**Supplementary Fig. 10** SPP suppression by a birefringent ETL with a reduced extraordinary refractive index $n_{\mathrm{e}}$. (a) TM power dissipation density versus the in-plane wavevector $k_x/k_0$ at $d_{\mathrm{ETL}}$ = 100 nm. (b) Evanescent loss (including the SPP mode, i.e., all near-field components beyond the light line of the organic layers) versus ETL thickness. The ordinary index of the ETL is fixed at 1.8; the stack is Ag 100 nm/ETL/EML 20 nm/HTL 50 nm/ITO 50 nm/$n_{\mathrm{sub}}$ = 1.8.

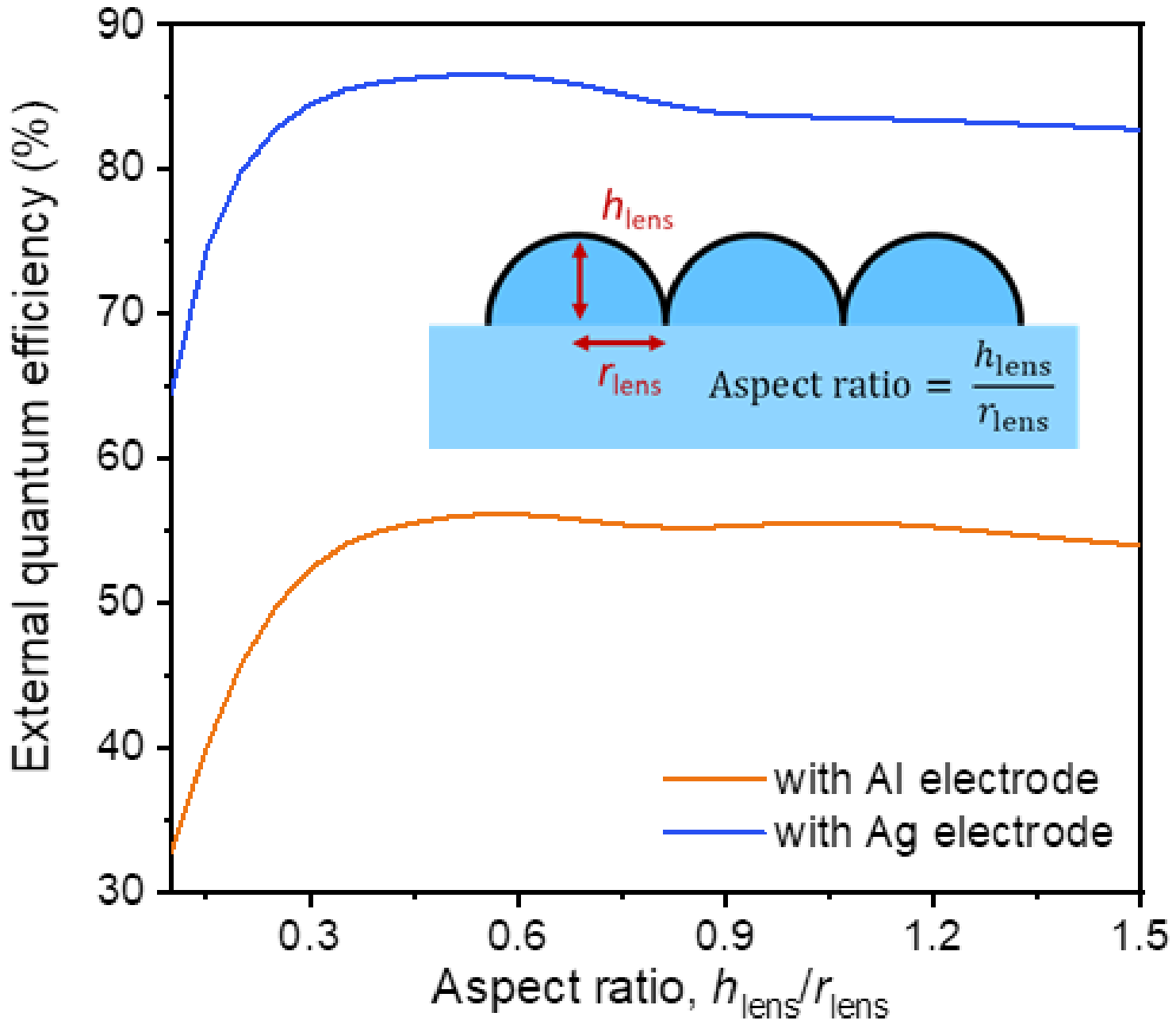

**Supplementary Fig. 11** EQE versus the aspect ratio (height/base radius) of a single lens in a hexagonally close-packed microlens array ($n_{sub} = n_{MLA} = 1.8$, Al and Ag reflectors). Without lenses, the planar substrate gives an EQE of 21.6% for Al and 25.5% for Ag. Computed with the angle-resolved model (Methods of the main text) at a single wavelength of 550 nm with PLQY = 1, $\Theta = 2/3$.

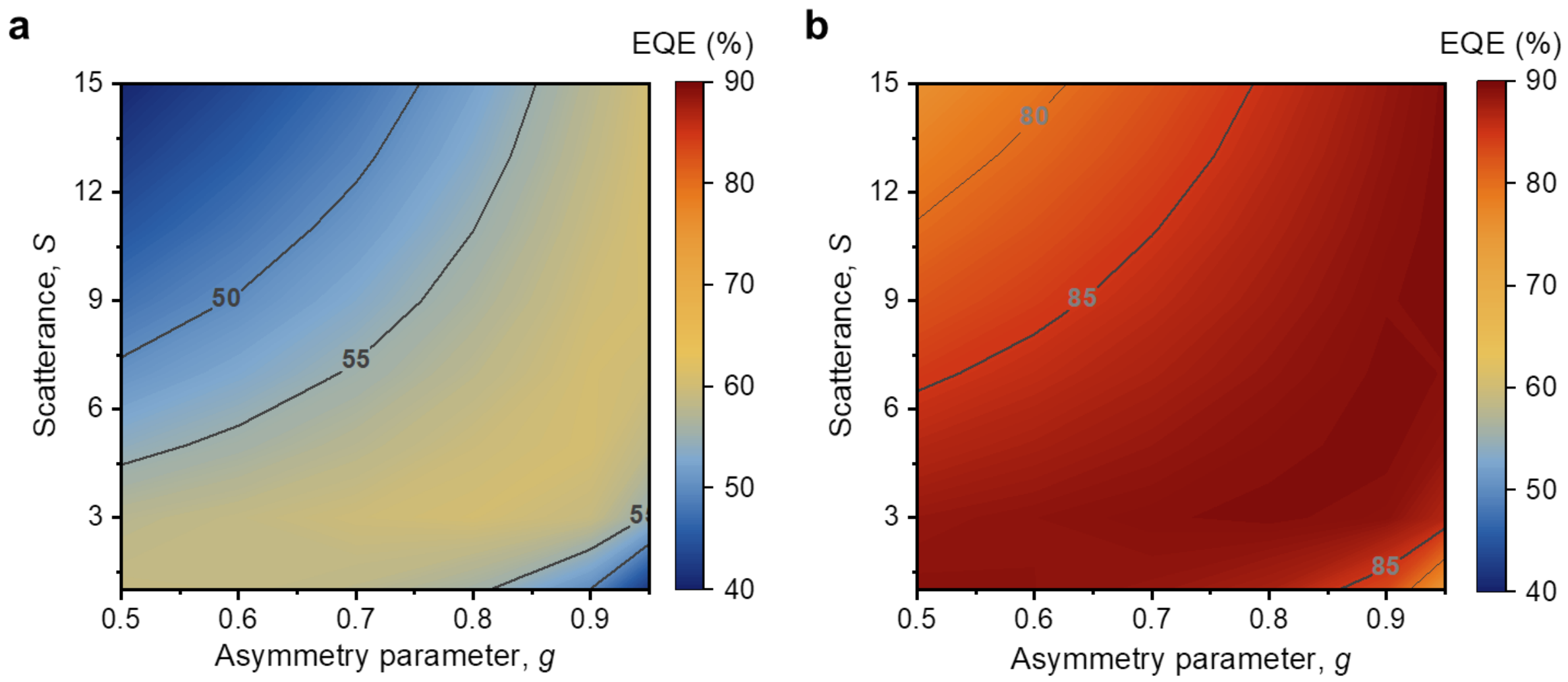

**Supplementary Fig. 12** EQE versus the scatterance $S$ and the asymmetry parameter $g$ of a scattering layer with (a) Al and (b) Ag reflector ($n_{\mathrm{sub}} = 1.8$). Computed with the angle-resolved model (Methods of the main text) with the 400–800 nm emission spectrum of Ir(ppy)$_2$acac, PLQY = 0.98, $\Theta = 0.76$.

**Supplementary Table 1. EQE of the near-unity design computed using measured optical constants of established materials and assuming PLQY = 1.** Stack: Ag 100 nm/B3PyMPM $d_{\mathrm{ETL}}$/TCTA:B3PyMPM 25 nm (emission zone at the center)/TAPC 180 nm/ITO 50 nm/substrate ($n_{\mathrm{sub}}$ = 1.8) with a hemispherical microlens array ($n_{\mathrm{MLA}}$ = 1.8, angle-resolved model). Measured optical constants (B3PyMPM: $n_{\mathrm{o}}$ = 1.821, $n_{\mathrm{e}}$ = 1.609, TCTA:B3PyMPM co-host: $n_{\mathrm{o}}$ = 1.833, $n_{\mathrm{e}}$ = 1.671, TAPC: $n_{\mathrm{o}}$ = 1.691, $n_{\mathrm{e}}$ = 1.664); for ITO, $n$ = 1.864 at 550 nm, with the wavelength-dependent optical constants of König et al.[7] and $k$ = 0.0032 at 550 nm, or with the extinction-coefficient spectrum scaled to 0.002 at 550 nm. $\Theta$ = 2/3 corresponds to an isotropic emitter. SPP loss refers to the entire evanescent component (including the SPP mode). Entries formatted as *x* / *y* are for the full emission spectrum of Ir(ppy)$_2$acac (400–700 nm, photon-number weighting; *x*) and single wavelength at 550 nm (*y*); the full-spectrum $\eta_{\mathrm{sta}}$ and EQE values are 0.6–0.9 percentage points below the 550-nm values, mostly because ITO and Ag absorb more strongly at shorter wavelengths. The $A'$ values shown are for a single wavelength of 550 nm. Equation (2) of the main text gives $\eta_{\mathrm{sta}}$ = 0.932–0.945 and EQE = 0.898–0.931, 1.6–3.2 percentage points above the angle-resolved model.

| $d_{\mathrm{ETL}}$ (nm) | $k_{\mathrm{ITO}}$ | $\Theta$ | $\eta_{\mathrm{sub}}$ (full/ @550 nm) | SPP loss | $A'$ | $\eta_{\mathrm{sta}}$ | EQE |
|---|---|---|---|---|---|---|---|
| 200 | 0.0032 | 2/3 | 0.962 / 0.963 | 0.023 / 0.023 | 0.029 | 0.904 / 0.912 | 0.870 / 0.878 |
| 200 | 0.0032 | 0.80 | 0.970 / 0.971 | 0.015 / 0.015 | 0.029 | 0.900 / 0.908 | 0.873 / 0.882 |
| 200 | 0.0032 | 0.90 | 0.975 / 0.976 | 0.010 / 0.010 | 0.029 | 0.897 / 0.906 | 0.875 / 0.884 |
| 200 | 0.002 | 2/3 | 0.965 / 0.966 | 0.023 / 0.023 | 0.024 | 0.921 / 0.929 | 0.889 / 0.897 |
| 200 | 0.002 | 0.80 | 0.973 / 0.973 | 0.015 / 0.015 | 0.024 | 0.918 / 0.926 | 0.893 / 0.902 |
| 200 | 0.002 | 0.90 | 0.978 / 0.979 | 0.010 / 0.010 | 0.024 | 0.916 / 0.924 | 0.896 / 0.905 |
| 300 | 0.0032 | 2/3 | 0.979 / 0.981 | 0.004 / 0.003 | 0.029 | 0.901 / 0.907 | 0.883 / 0.890 |
| 300 | 0.0032 | 0.80 | 0.980 / 0.981 | 0.003 / 0.002 | 0.029 | 0.896 / 0.903 | 0.879 / 0.886 |
| 300 | 0.0032 | 0.90 | 0.981 / 0.982 | 0.002 / 0.001 | 0.029 | 0.893 / 0.900 | 0.876 / 0.883 |
| 300 | 0.002 | 2/3 | 0.983 / 0.984 | 0.004 / 0.003 | 0.024 | 0.921 / 0.927 | 0.905 / 0.912 |
| 300 | 0.002 | 0.80 | 0.984 / 0.984 | 0.003 / 0.002 | 0.024 | 0.918 / 0.924 | 0.903 / 0.909 |
| 300 | 0.002 | 0.90 | 0.984 / 0.985 | 0.002 / 0.001 | 0.024 | 0.915 / 0.921 | 0.901 / 0.907 |

**Supplementary References**